\documentclass[journal]{IEEEtran}
\usepackage{graphicx}
\usepackage{cite}
\usepackage{color}
\usepackage[hidelinks]{hyperref}
\usepackage{tikz}

\makeatletter
\let\MYcaption\@makecaption
\makeatother

\usepackage[font=footnotesize]{subcaption}

\makeatletter
\let\@makecaption\MYcaption
\makeatother

\usepackage{amsmath,amsfonts,amssymb,mathrsfs} 
\usepackage{array}

\usepackage[full]{bib_sty}

\graphicspath{{images_final/}}

\DeclareMathOperator*{\argmin}{arg\,min}
\newcommand*{\tran}{^{\mkern-0.5mu\mathsf{T}}}

\usepackage{mathtools}
\usepackage{cases}

\usepackage[normalem]{ulem}

\newtheorem{myalg}{Algorithm}

  \definecolor{tuewarmred}{rgb}{0.969,0.192,0.192}
  \definecolor{tueprocesscyan}{rgb}{0.000,0.635,0.871}
  \definecolor{tuecyan}{rgb}{0.000,0.635,0.871}
  \definecolor{tuered}{rgb}{0.839,0.000,0.290}
  \definecolor{tueblue}{rgb}{0.000,0.400,0.800}
  \definecolor{tuedarkblue}{rgb}{0.063,0.063,0.451}
  \definecolor{tueorange}{rgb}{1.000,0.604,0.000}
  \definecolor{tueyellow}{rgb}{1.000,0.867,0.000}
  \definecolor{tuelightgreen}{rgb}{0.518,0.824,0.000}
  \definecolor{tuegreen}{rgb}{0.000,0.675,0.510}

\begin{document}

\title{\Huge Identifying Position-Dependent Mechanical Systems: A Modal Approach Applied to a Flexible Wafer Stage}
\author{
	\vskip 1em
	{\color{black}
	Robbert Voorhoeve,
	Robin de Rozario,
	Wouter Aangenent,
	and Tom Oomen, \emph{Senior Member, IEEE}
	}

	\thanks{
		
		{\color{black}
		This research is supported by ASML Research as part of the TU/e Impulse I program and is part of the research program VIDI with project number 15698, financed by the Netherlands Organization for Scientific Research (NWO).
		
		Robbert Voorhoeve, Robin de Rozario, and Tom Oomen are with the Eindhoven University of Technology, Department of Mechanical Engineering, Control Systems Technology group, Eindhoven, The Netherlands (e-mail addresses: r.j.voorhoeve{@}gmail.com, r.d.rozario{@}tue.nl, t.a.e.oomen{@}tue.nl). 
		
		Wouter Aangenent is with ASML Research Mechatronics, Veldhoven, The Netherlands (e-mail address: wouter.aangenent{@}asml.com).
		
		This article has been accepted for publication in IEEE Transactions on Control Systems Technology (DOI: 10.1109/TCST.2020.2974140).
		
		\textcopyright 2020 IEEE. Personal use of this material is permitted. Permission from IEEE must be obtained for all other uses, in any current or future media, including reprinting/republishing this material for advertising or promotional purposes, creating new collective works, for resale or redistribution to servers or lists, or reuse of any copyrighted component of this work in other works.
				}
	}
}

\maketitle
	
\begin{abstract}
Increasingly stringent performance requirements for motion control necessitate the use of increasingly detailed models of the system behavior. Motion systems inherently move, therefore, spatio-temporal models of the flexible dynamics are essential. In this paper, a two-step approach for the identification of the spatio-temporal behavior of mechanical systems is developed and applied to a lightweight prototype industrial wafer stage. The proposed approach exploits a modal modeling framework and combines recently developed powerful linear time invariant (LTI) identification tools with a spline-based mode-shape interpolation approach to estimate the spatial system behavior. The experimental results for the wafer stage application confirm the suitability of the proposed approach for the identification of complex position-dependent mechanical systems, and its potential for motion control performance improvements.
\end{abstract}


{}

\definecolor{limegreen}{rgb}{0.2, 0.8, 0.2}
\definecolor{forestgreen}{rgb}{0.13, 0.55, 0.13}
\definecolor{greenhtml}{rgb}{0.0, 0.5, 0.0}

\section{Introduction}

Increasingly stringent performance requirements for precision motion systems lead to a situation where the flexible dynamics of moving machine components need to be actively modeled and controlled. Typical examples include the wafer stages in lithographic wafer scanners \cite{Vandewal2002,Oomen2014}. Traditionally, these stages can be accurately approximated as a rigid body in the frequency range relevant for control \cite{MunnigSchmidt2011,Oomen2018}, thereby enabling static decoupling of the rigid-body dynamics and subsequent decentralized control design \cite{Fleming2014}. Furthermore, when this rigid-body approximation is used, the spatial system behavior follows directly from the stage geometry. Due to increasing accuracy and speed requirements, the flexible dynamics  in future systems can no longer be neglected and need to be explicitly addressed. Approaches to address the resulting complex spatio-temporal system behavior include over-actuation and over-sensing \cite{Herpen2014}, spatial vibration control \cite{moheimani2003}, multivariable robust control \cite{Vandewal2002}, and inferential control of unmeasured performance variables \cite{Oomen2015}. Invariably, these approaches are characterized by an increased reliance on model-based control design procedures, necessitating the development of control-relevant, efficient, and numerically reliable identification algorithms capable of dealing with the complex spatio-temporal system behavior \cite{Oomen2014,Herpen2014aut,Voorhoeve2016}.

\begin{figure}[t]
\begin{center}
  \includegraphics[width=.91\linewidth]{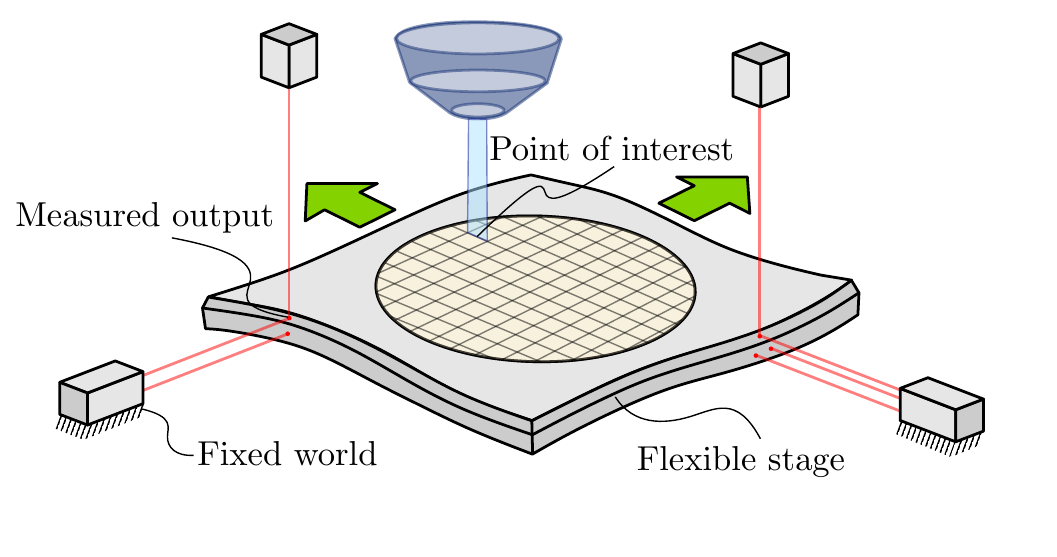}
  \caption{Schematic flexible wafer-stage system.} 
\label{fig:wafer_schem}
\end{center}
\end{figure} 

The flexible dynamics of these systems in conjunction with the fact that these motion systems move lead to position-dependent system behavior \cite{Oomen2018,daSilva2008}. In this paper, mechanical systems consisting of a single flexible moving body are considered and deformations are assumed to be small. As an example, consider the schematic flexible wafer-stage system in Figure \ref{fig:wafer_schem}. Here, the flexible wafer-stage moves in relation to the sensors, which are connected to the fixed world. As a result of this relative motion, the sensors measure the position at different points on the flexible structure, and therefore the spatial behavior of this flexible system is observed differently depending on the relative location of the sensors. However, due to the fact that there is only a single moving body and because deformations are small, the structural dynamics of the flexible body do not change as the system moves. Still, as a result of the position dependency in the measurement system, the system dynamics as observed by the sensors are no longer Linear-Time-Invariant (LTI), necessitating a deviation from the standard LTI control design techniques used in the context of high-precision motion systems. 

Even though standard linear control design approaches are no longer sufficient to control the position dependent dynamics, the particular properties of the considered class of systems can be exploited to utilize control approaches that remain relatively close to LTI control theory. Approaches in literature that utilize additional system information to retain some aspects of LTI control theory include the gain-scheduling control approach \cite{Shamma1992,Rugh2000,Leith2000} and the Linear Parameter-Varying (LPV) control framework \cite{Packard1994,Apkarian1995,Scherer2001,Hoffmann2015}, which formalizes the gain-scheduling method by ensuring stability and performance through a rigorous model-based mathematical approach. While the class of systems considered in this paper can be readily treated within the LPV control framework, the present paper proposes an approach that explicitly utilizes the additional assumptions of  a single moving body to simplify the problem.

A key challenge for systematic LPV control is the availability of accurate LPV models. The need for accurate LPV models spurred the development of LPV system identification with a strong focus on black-box parametric models \cite{Bamieh2002,Lovera2007,VanWingerden2009,Toth2010}. This resulted in a well-developed theoretical framework that categorizes the identification techniques in local \cite{DeCaigny2011,Vizer2013}, and global approaches, e.g., \cite{Felici2007,Goos2016}. Depending on the application, both approaches have been reported to effectively support the identification of practically relevant systems \cite{Turk2018}, albeit that the identification of systems with high dynamic order remains challenging \cite{VanWingerden2009,Cox2018}. Due to the high model complexity associated with general LPV modeling, the success of black-box LPV approaches is limited for the identification of mechanical systems with a large number of resonant modes \cite{GrootWassink2005,Steinbuch2003}, showing a need to reduce the modeling complexity by using additional prior system knowledge. 

Although many important developments have progressed LPV identification for control, the continuously increasing complexity of motion systems necessitates a practical identification approach of reduced complexity that is systematic, accurate, and user-friendly. A key step in this paper is to utilize the knowledge that the system consists of a single moving body with small deformations to derive a parsimonious model-set, significantly reducing the modeling complexity compared to a full LPV approach. Furthermore, recently developed efficient and reliable LTI identification tools are employed to obtain accurate and coherent local models which are particularly suited to a subsequent interpolation step to obtain the desired position-dependent model. The aim of this paper is to develop an effective and practical approach for the identification of position-dependent mechanical systems.

The main contributions of this paper are the following.
\begin{enumerate}
\item A two-step modal identification framework for position-dependent mechanical systems, including
\begin{enumerate}
\item a flexible framework of parameterizations and algorithms aimed at obtaining accurate modal LTI models of complex mechanical systems,
\item an approach for the interpolation of identified mode-shapes to obtain position-dependent models for control of flexible mechanical systems.
\end{enumerate}
\item Application and validation of the developed approach on a state-of-the-art industrial wafer stage setup.
\end{enumerate}
These contributions are inextricably linked as the proposed two-step position-dependent identification framework is explicitly developed with the goal of being able to handle the complexity of such a state-of-the-art industrial system. The novelty of this paper therefore lies in the formulation and validation of the full position-dependent identification framework, which, while making use of  previous results, including but not limited to previous results by the authors, e.g., in \cite{Voorhoeve2016,Rozario2017}, has not been previously published.

The identification methods used in this paper have parallels with the field of experimental modal analysis. Research in this field has seen significant developments in the past decade \cite{guillaume2003,Cauberghe2004,Vayssettes2014}, in part due to recent consolidation efforts of the approaches used in modal analysis and system identification \cite{fleming2003,Pintelon2007,Reynders2012}. Contrary to modal analysis, the modeling goal in this paper is to obtain position-dependent models for control. This is reflected by the emphasis on accurate mode-shape interpolation and the possibility to incorporate control-relevant identification criteria as in, e.g., \cite{Oomen2014}.

The outline of this paper is as follows. In Section \ref{Sect:prob_form_OAT}, the experimental setup is introduced and the control challenges as well as the position dependent modeling problem for this setup are formulated. In Section \ref{Sect:ID_approach}, the proposed two-step position-dependent identification approach is explained. In Section \ref{Sect:LTI_ID}, the LTI identification approach and the obtained results for the prototype wafer stage system are presented. In Section \ref{Sect:interp_OAT}, the approach and results for the mode shape interpolation are presented. In Section \ref{Sect:Control_app}, a discussion is presented on the applicability of the proposed approach for control. In Section \ref{Sect:Conclusions_n_out} the conclusions of this paper are formulated as well as an outlook on ongoing research.

\begin{figure}[t]
\begin{center}
\includegraphics[width=.75\linewidth]{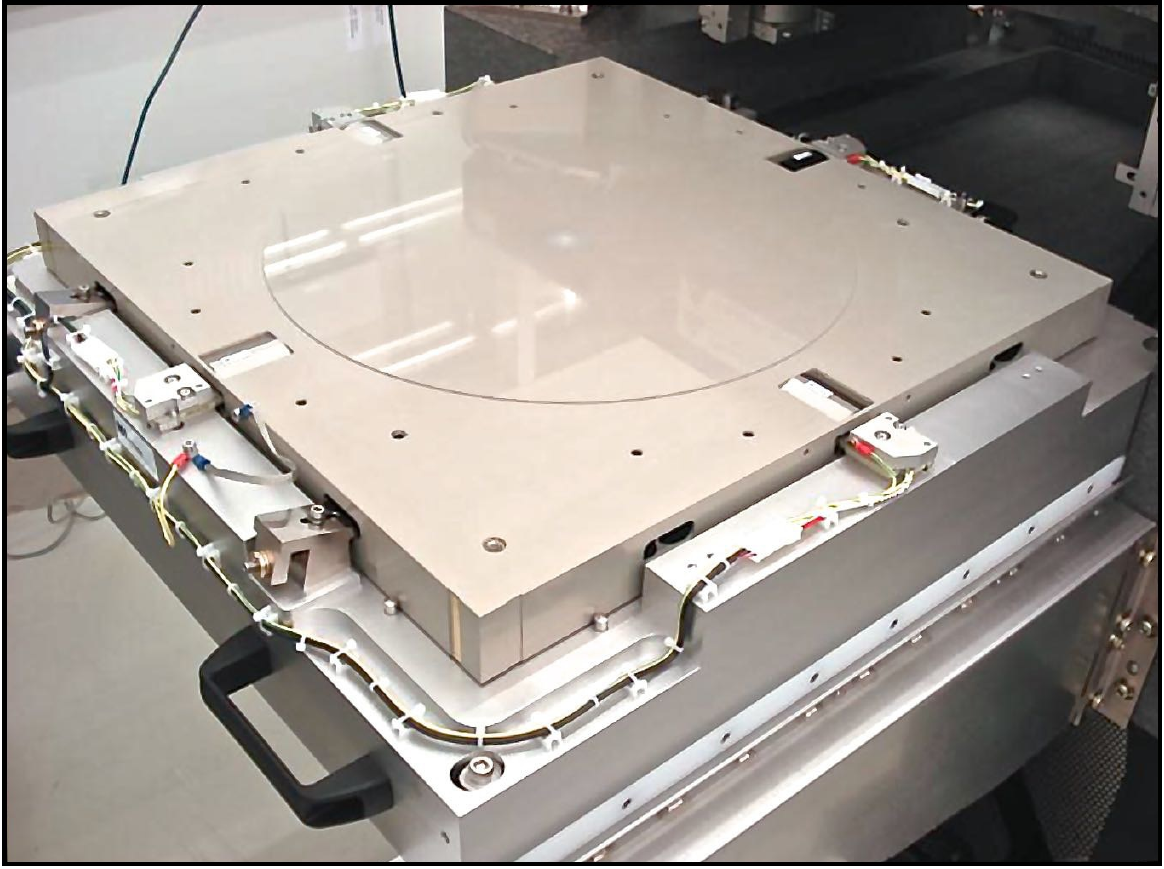}
\caption{The experimental wafer-stage setup.} 
\label{fig:OAT}
\end{center}
\end{figure}

\section{Experimental Setup and Problem Formulation}
\label{Sect:prob_form_OAT}

In this section the experimental wafer-stage setup and related control challenges are outlined. In Section \ref{Sect:Pos_dep_mod}, the position-dependent modeling problem is formulated, as is considered in this paper.

\subsection{Experimental Setup}\label{Sect:exp_set_OAT}
The experimental setup considered in this paper is the Over-Actuated-Testrig (OAT), as is shown in Figure \ref{fig:OAT}. The system is considered here as an analogue to the flexible wafer-stage as depicted in Figure \ref{fig:wafer_schem}. The system is controlled in six motion degrees of freedom and is magnetically levitated having no mechanical connection to the fixed world. Furthermore, the experimental system is equipped with additional actuators and sensors to facilitate the spatio-temporal identification of the flexible dynamics. While these additional actuators and sensors are available on the experimental setup, they are not available in the considered wafer-stage system as depicted in Figure \ref{fig:wafer_schem} for which the experimental system is an analogue. Therefore, these additional actuators and sensors are only used for identification and are considered to be unavailable for the control of the system. In this paper, only the out-of-plane motions are considered, i.e., the motions perpendicular to the surface of the wafer-stage, both for visualization purposes and due to the availability of multiple spatially distributed sensors in this direction.

\begin{figure}[t]
\begin{center}
\includegraphics[width=.8\linewidth]{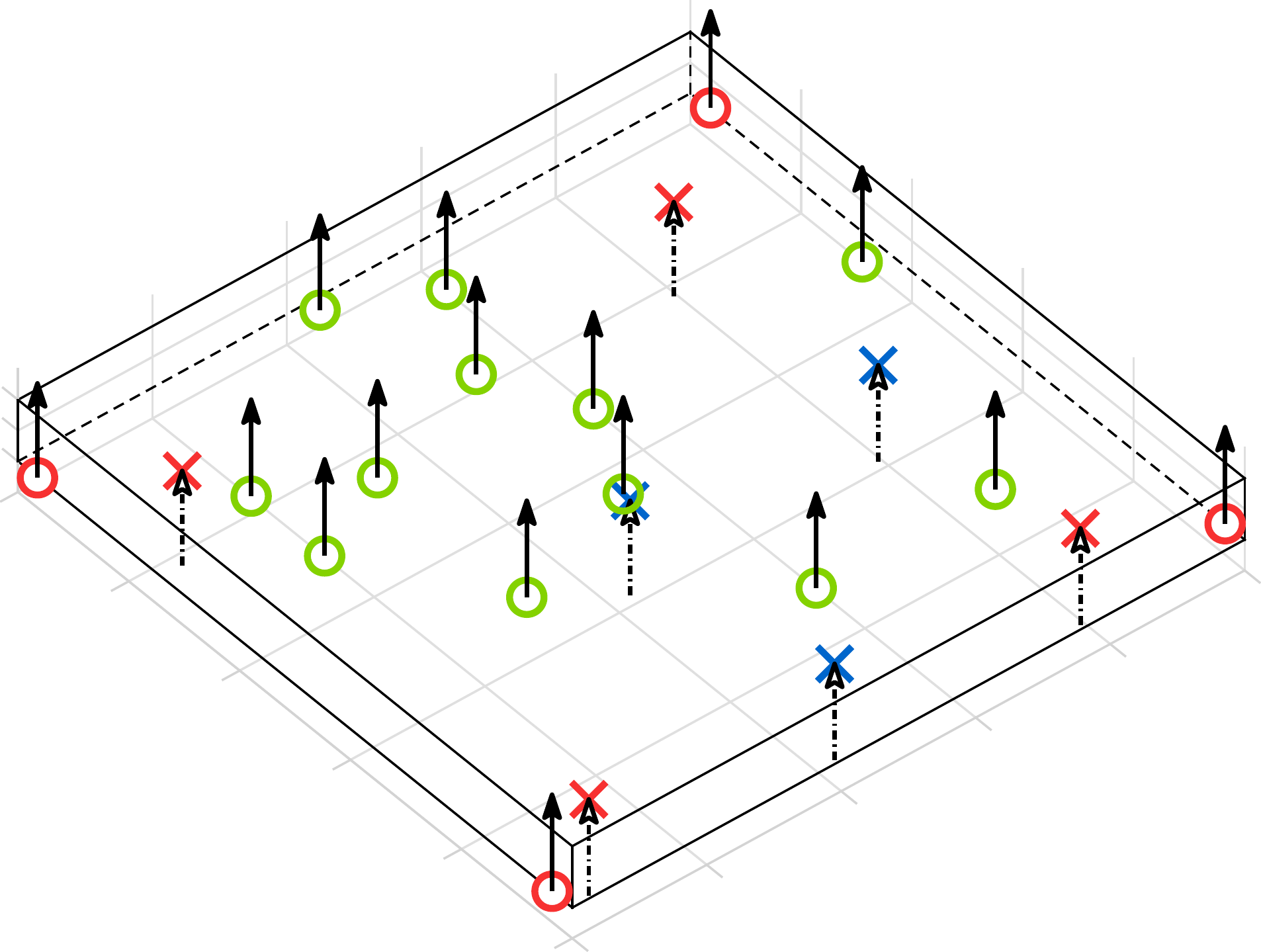}
\caption{Actuator and sensor configuration of the considered experimental setup. The locations of actuators used for control and identification are marked with red crosses (\protect\tikz[baseline=-0.6ex,x=1pt,y=1pt]{\protect\draw[tuewarmred,very thick] [-] (0,-2.75) -- (5.5,2.75);\protect\draw[tuewarmred,very thick] [-] (0,2.75) -- (5.5,-2.75);}) actuators used only for identification as blue crosses (\protect\tikz[baseline=-0.6ex,x=1pt,y=1pt]{\protect\draw[tueblue,very thick] [-] (0,-2.75) -- (5.5,2.75);\protect\draw[tueblue,very thick] [-] (0,2.75) -- (5.5,-2.75);}), sensors used for identification and control as red circles (\protect\tikz[baseline=-0.6ex,x=1pt,y=1pt]{\protect\draw[tuewarmred,very thick] (7,0) circle (2.75);}), and sensors used only for identification as green circles (\protect\tikz[baseline=-0.6ex,x=1pt,y=1pt]{\protect\draw[tuelightgreen,very thick] (7,0) circle (2.75);}).}
\label{fig:OAT_io}
\end{center}
\end{figure}

The out-of-plane sensors and actuators used for the experiments in this paper are shown in Figure \ref{fig:OAT_io}. Seven actuators, depicted by crosses in Figure \ref{fig:OAT_io}, four of which, shown in red, are used for closed-loop control and the remaining three, shown in blue, are used to apply additional spatially distributed excitation for identification. Sixteen sensors are available for identification, as shown in Figure \ref{fig:OAT_io} by circles. Similarly, a distinction is made between the sensors used for closed-loop control, shown in red, and those only used for the spatially distributed identification, which are shown in green. 

\subsection{Control Challenges}
Consider the schematic wafer-stage setup shown in Figure \ref{fig:wafer_schem}. For this setup, two key control challenges are recognized which are related to the position-dependent system behavior. First, as previously outlined, the relative motion of the wafer stage with respect to the sensors leads to an input-output behavior that is no longer LTI, necessitating a position/dependent modeling and control perspective. Second, the point of interest, i.e., the point on the wafer that is being exposed in the photo-lithographic process, also changes as the wafer stage moves. This involves the control of an unmeasured performance variable since the point of interest is not directly measured. This results in a position-dependent inferential control problem. See, for example, \cite{Oomen2015,Voorhoeve2016a,Rozario2017}, for control design approaches for such problems.

\begin{figure}[t]
\begin{center}
  \includegraphics[height=3.5cm]{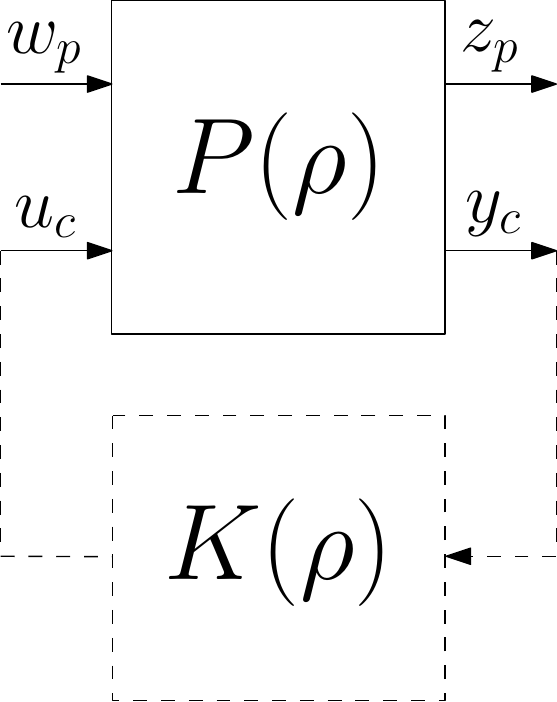}
  \caption{The LPV standard plant framework.}
  \label{fig:wafer_contr}
\end{center}
\end{figure}

The LPV standard plant framework, as depicted in Figure \ref{fig:wafer_contr}, can be used to describe both these control problems. Here, $u_c$ and $y_c$ are the output and input signals available for control, $w_p$ is the generalized disturbance signal and $z_p$ is the generalized performance signal, which in this case involves the positioning error of the point of interest. This control problem, with the LPV standard plant $P(\rho)$ in a generalized feedback interconnection with an LPV controller $K(\rho)$, has been studied extensively in, e.g., \cite{Packard1994,Apkarian1995,Scherer2001} and, for appropriately bounded sets of scheduling variable trajectories, i.e., $\rho(t) \in \mathscr{D}$, efficient algorithms exist for various robust and optimal control problems defined in this framework.

The main difficulty for the practical application of these methods concerns the availability of an accurate LPV system-model $G(\rho)$, which is the part of the standard plant $P(\rho)$ pertaining to the physical system that is to be controlled. That this is a difficult problem is evidenced by the fact that accurate modeling of LTI precision systems is already considered to be a challenging problem, see, e.g., \cite{Herpen2014aut,Voorhoeve2016}. Accurate modeling for LPV systems is generally significantly more challenging, since the incorporation of a scheduling parameter dependency typically leads to a highly increased model complexity \cite{VanWingerden2009}. Due to the additional constraints on the considered class of systems, consisting of one flexible moving body and assuming small deformations, the complexity of the problem can be significantly reduces. This problem of obtaining accurate position-dependent models for such mechanical systems, such as the wafer-stage example considered here, is addressed in the remainder of this paper.

\subsection{Position-Dependent Modeling Problem}
\label{Sect:Pos_dep_mod}

In the considered wafer-stage example, the scheduling variables $\rho$ relate to the position of the wafer stage. 
Due to the fact that there is only a single moving body, the structural dynamics of the flexible body are invariant to the position of the wafer stage with respect to the sensors. Furthermore, deformations are considered to be small in the sense that the structure does not deform in such a way that it influences the dynamics of the systems. As a result, the $A$ matrix of the state-space description for this class of systems does not depend on the scheduling variable $\rho$, see, e.g., \cite{Gawronski1998,Preumont2018}. Furthermore, for the considered system the positions of the actuators are fixed relative to the wafer-table and therefore the input matrix  $B$ is also not position dependent. Therefore, the scheduling variables have no influence on the system states and as a result there is no memory in the system pertaining to the past-trajectory of the scheduling variables. The system is therefore only dependent on the current, instantaneous, value of the scheduling parameters. In state-space the considered system-model is described as,

\begin{numcases}{G(\rho):}
\dot{x}(t) = A x(t) + B u(t) \label{eq:LPV_1}\\
y(t) = C_y(\rho(t)) x(t) + D_{y}(\rho(t)) u(t)\label{eq:LPV_2}\\
z(t) = C_z(\rho(t)) x(t) + D_{z}(\rho(t)) u(t)\label{eq:LPV_3}.
\end{numcases}

In the wafer-stage example, both the system outputs $y(t)$ and the performance variables $z(t)$, i.e., the point of interest position, can be considered as specific local instances of the out-of-plane deflection of the surface of the wafer stage. In this interpretation, the relevant output for this system is this out-of-plane deflection, which is denoted here as $\mathfrak{z}(\varrho,t)$, where $\varrho$ is the in-plane coordinate of the point for which the deflection is considered. The modeling problem is then to identify from experimental data of the system, a model for the system behavior of the form
\begin{numcases}{G(\varrho):}
\dot{x}(t) = A x(t) + B u(t)\label{eq:spa_temp_1}\\
\mathfrak{z}(\varrho,t) = \mathfrak{C}(\varrho) x(t) + \mathfrak{D}(\varrho) u(t)\label{eq:spa_temp_2}\,.
\end{numcases}
The model in the form \eqref{eq:LPV_1}--\eqref{eq:LPV_3} is recovered from this description by including the static geometric relations between the position of the wafer stage $\rho$ and the coordinates $\varrho$, at which the sensors view the wafer-stage as well as the coordinate of the point of interest. This is a simple affine relation dependent on the definitions of the origins for two coordinate systems and the sensor locations, i.e., for a certain sensor $y_1$, $C_{y_1}(\rho(t)) = \mathfrak{C}(\varrho_{y_1}(\rho(t))$ with $\varrho_{y_1}(\rho(t)) = \varrho_{{y_1},0} + \rho(t)$.  The modeling problem considered in the remainder of this paper is therefore to model the system of the form \eqref{eq:spa_temp_1}--\eqref{eq:spa_temp_2}.

\section{Position-Dependent Modeling Approach}
\label{Sect:ID_approach}

In this section the proposed position-dependent modeling approach is described. First, the modal modeling framework for mechanical systems is outlined. Next, the proposed two-step identification approach is developed.

\subsection{Modal Models of Mechanical Systems}
\label{Sect:Modal_models}
The quantity of interest is the out-of-plane deflection, $\mathfrak{z}(\varrho,t): \mathfrak{D} \times T \mapsto \mathbb{R}$, of the surface of a flexible body, which is modeled here as a continuum. The domain $\mathfrak{D} \in \mathbb{R}^2$ of the coordinate $\varrho$ is the surface of the considered structure, e.g., the $(x,\,y)$ surface of the wafer-stage. To model the spatio-temporal evolution of $\mathfrak{z}(\varrho,t)$, a basis-function expansion is used for time--space separation, see, e.g., \cite{Li2010}, i.e.,
\begin{equation}
\mathfrak{z}(\varrho,t) = \sum_{i=1}^{n_q} w_i(\varrho) q_i(t)\,.\label{eq:basis_exp}
\end{equation}
For $n_q \rightarrow \infty$ this expansion converges as long as $\{w_i(\varrho)\}_{i=1}^\infty$ is a convergent set of functions for the class of continuous functions on the spatial domain $\mathfrak{D}$ \cite{Li2010}. A widely used method that is applicable for any geometrically complex domain $\mathfrak{D}$ is the Finite Element Method (FEM). This approach uses many localized basis functions to accurately approximate the spatial system behavior \cite{Li2010}.

The temporal system behavior for this basis function approach is determined by the dynamics of the generalized coordinates, $q(t) = [q_1(t) \; \ldots \; q_{n_q}(t)]\tran$. Under the assumption of small rotations and strains, and assuming that the material is linear elastic obeying Hooke's Law, the equations of motion that govern the temporal input-output behavior of a mechanical system are given by the set of coupled second order ordinary differential equations \cite[Section 2.2]{Gawronski1998}
\begin{align}
\mathcal{M} \ddot{q}(t) + \mathcal{D}\dot{q}(t) + \mathcal{K} q(t) = Q u(t)\,, \label{eq:mech_temp} 
\end{align}
where $\mathcal{M} \in \mathbb{R}^{n_q \times n_q}$ is the mass matrix, $\mathcal{D} \in \mathbb{R}^{n_q \times n_q}$ the damping matrix, $\mathcal{K} \in \mathbb{R}^{n_q \times n_q}$ the stiffness matrix, and $Q\in \mathbb{R}^{n_q \times n_u}$ the input distribution matrix.

The set of coupled equations of motion \eqref{eq:mech_temp} can be decoupled for the undamped case by transforming to a modal description, which is obtained by solving the generalized eigenvalue problem
\begin{equation}
\left[\mathcal{K}-\omega_i^2 \mathcal{M}\right] \phi_i = 0\,,\qquad\qquad i=1,\ldots,\,n_q\,.
\end{equation}
The eigenvalues, $\omega_i^2$, are the squared undamped resonance-frequencies of the modes, and the eigenvectors, $\phi_i$, are the associated mode shapes as parameterized in the basis $W(\varrho) = [w_1(\varrho) \;\ldots \; w_{n_q}(\varrho)]$. By applying the substitution $q =\Phi \eta$, where $\Phi = [\phi_1 \;\ldots \; \phi_{n_q}]$, and multiplying \eqref{eq:mech_temp} with $\Phi^{-1}\mathcal{M}^{-1}$ yields
\begin{numcases}{G_m(\varrho):}
I\ddot{\eta}(t) + \mathcal{D}_m\dot{\eta}(t) + \Omega^2 \eta(t) = \mathcal{R}u(t)\,, \label{eq:modal_temp}\\
\mathfrak{z}(\varrho,t) = \mathcal{L}(\varrho)\eta(t)\label{eq:modal_spat}\,,
\end{numcases}
where $\mathcal{D}_m = \Phi^{-1}\mathcal{M}^{-1} \mathcal{D} \Phi$, $\Omega^2 = \mathrm{diag}([\omega_1^2 \;\ldots \; \omega_{n_q}^2])$, $\mathcal{L}(\varrho) = W(\varrho) \Phi$, and $\mathcal{R} = \Phi^{-1}\mathcal{M}^{-1} Q$.

In the context of identification for control, low-order models are desired that are accurate in a limited frequency band of interest \cite{Oomen2014}. This means that only a limited number of modes, $n_m < n_q$, are required to model the relevant temporal system behavior \cite{Gawronski1998}. Modeling the spatial system behavior, using a generic set of basis functions $\mathcal{L}(\varrho) = W(\varrho)\Phi$, typically requires a large number of basis functions leading to a high modeling complexity. In this paper, a two-step identification approach is proposed to directly identify the mode-shapes, i.e., the columns of $\mathcal{L}(\varrho)$ from measured data.

\subsection{Two-Step Identification Approach}
To identify the spatio-temporal system behavior, measurement data is first obtained in experiments with fixed sensor locations $\varrho_i$. As a result of the fixed sensor locations the input-output system behavior is linear time invariant, similar to the local approach in LPV identification. Experiments are performed with various fixed sensor locations covering the domain $\mathfrak{D}$. An LTI system model is then identified from the obtained experimental data where the model is parameterized in modal form, i.e., \eqref{eq:modal_temp}--\eqref{eq:modal_spat}, with $n_m$ modes. Instead of the position-dependent output equation \eqref{eq:modal_spat} the measured, spatially sampled, outputs $\mathfrak{z}_s(t)$ are modeled as
\begin{equation}
\mathfrak{z}_s(t) = \begin{bmatrix}
\mathfrak{z}({\varrho}_1,t) \\ \vdots \\ \mathfrak{z}({\varrho}_{n_\varrho},t)
\end{bmatrix} = {\mathcal{L}_s}\eta(t),\quad\; {\mathcal{L}}_s \approx \begin{bmatrix}
\mathcal{L}({\varrho}_1) \\ \vdots \\ \mathcal{L}({\varrho}_{n_\varrho})
\end{bmatrix}\label{eq:frozen_z}
\end{equation}
where the parameters in ${\mathcal{L}_s} \in \mathbb{R}^{n_\varrho \times n_m}$ are considered as spatially sampled estimates of the mode shapes $\mathcal{L}(\varrho)$.

This first step requires the LTI identification of a complex mechanical system with a high model order and many inputs and outputs. The identification of such complex mechanical systems requires the use of efficient and numerically reliable identification approaches, as have been developed and investigated in, e.g., \cite{Herpen2014aut,Voorhoeve2016}.

In the second step, the spatial mode shapes ${\mathcal{L}}(\varrho)$ are estimated from the identified parameters in ${\mathcal{L}}_s$. In this step, interpolation techniques are used to reconstruct continuous mode shapes based on these spatially sampled estimates. Since this step involves the interpolation of spatial functions in $\varrho$ and not of systems that dynamically depended on a scheduling variable $\rho$, the interpolation pitfalls as shown in \cite{Toth2007} are avoided. In Section \ref{Sect:interp_OAT}, a promising robust and physically motivated interpolation approach is proposed, but several other interpolation techniques, which might be more suitable for other applications, can be used in this second step of the proposed two-step approach.

In summary, the proposed two-step approach aims to:
\begin{enumerate}
\item Identify the modal mechanical LTI model given by \eqref{eq:modal_temp} and \eqref{eq:frozen_z}, i.e., estimate the parameters in ${\mathcal{L}}_s,$ $\Omega^2,$ $\mathcal{D}_m,$ $\mathcal{R}$.
\item Estimate the mode-shapes ${\mathcal{L}}(\varrho)$ based on the spatially sampled mode-shapes ${\mathcal{L}}_s$
\end{enumerate}

\section{LTI Identification of Spatially Sampled Systems}
In this section, the first step of the proposed identification approach is outlined, which is the LTI identification of the spatially sampled system $G_s$.

\label{Sect:LTI_ID}
\subsection{Methods}
The LTI identification approach considered here involves a number of key aspects. First, a non-parametric identification approach is considered, aimed at obtaining accurate Frequency Response Function (FRF) estimates of the spatially sampled system $G_s$. Second, the modal parametrization as used in this paper is defined. Third, a black-box Matrix Fraction Description (MFD) parametrization is employed, which is parameterized such that it closely matches the modal parametrization. Fourth, the identification algorithms used to estimate the system models from the measured data are explained.

\subsubsection{Non-Parametric Identification}
\begin{figure}[t]
\begin{center}
\includegraphics[width=.7\linewidth]{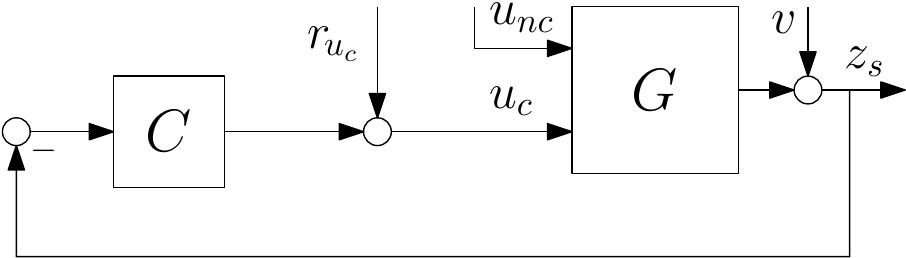}
\caption{Closed-loop identification scheme.}
\label{fig:CL_ID}
\end{center}
\end{figure}

The non-parametric frequency response function for the wafer-stage system is estimated using the robust multisine approach as explained in, e.g., \cite[Section 3.7]{PintelonSchoukens12}. The rigid-body motions of the system need to be controlled for stable operation, therefore all experiments are performed in a closed-loop configuration. The closed-loop identification scheme is shown in Figure \ref{fig:CL_ID}. A distinction is made between inputs which are used in the control loop $u_c(t)$ and those that are not used in the control loop $u_{nc}(t)$. The control inputs $u_c(t)$ also include the in-plane actuator signals that are used to stabilize the in-plane rigid-body modes. The excitation signals used for system identification are the non-control inputs $u_{nc}(t)$ and the additive perturbation signals $r_{u_c}(t)$ as shown in Figure \ref{fig:CL_ID}. The applied excitation signals are all random-phase multisines with a flat amplitude spectrum which are successively applied to the single inputs in separate experiments.

With a total of $16$ out-of-plane sensors, $8$ control inputs, including $4$ in the in-plane direction, and $3$ non-control inputs, the identification problem involves first identifying a $24 \times 11$ closed-loop FRF given by
\begin{equation}
\tilde{P}_{CL}(\Omega_k) = \begin{bmatrix}
\tilde{P}_{z_s\leftarrow r_{u_c}}(\Omega_k) & \tilde{P}_{z_s\leftarrow u_{nc}}(\Omega_k)\\\tilde{P}_{u_c \leftarrow r_{u_c}}(\Omega_k) & \tilde{P}_{u_c \leftarrow u_{nc}}(\Omega_k)
\end{bmatrix}\,, \label{eq:CL_id_prob}
\end{equation}
where the notation $\tilde{P}_{y\leftarrow x}(\Omega_k)$ is used to denote the identified empirical transfer function estimate (ETFE) from the input signal $x$ to output signal $y$ at frequency point $\Omega_k$. To obtain the FRF of the open-loop system $G$ from this closed-loop FRF the following relation is used,
\begin{equation}
\tilde{G}(\Omega_k) = \begin{bmatrix}
\tilde{P}_{z_s\leftarrow r_{u_c}} & \tilde{P}_{z_s\leftarrow u_{nc}}
\end{bmatrix}  \begin{bmatrix}
\tilde{P}_{u_c \leftarrow r_{u_c}} & \tilde{P}_{u_c \leftarrow u_{nc}}\\
\tilde{P}_{u_{nc} \leftarrow r_{u_c}} & \tilde{P}_{u_{nc} \leftarrow u_{nc}}
\end{bmatrix}^{-1}\,,\label{eq:CL_to_OL}
\end{equation}
where the arguments $\Omega_k$ have been omitted on the right hand side of this expression for brevity. In \eqref{eq:CL_to_OL} $\tilde{P}_{u_{nc} \leftarrow r_{u_c}} = \mathbf{0}$ and $\tilde{P}_{u_{nc} \leftarrow u_{nc}} = I$, see, e.g., \cite[Appendix A]{Oomen2014} for additional detail on this closed-loop identification approach. By removing the in-plane input directions, the $16\times 7$ FRF of the system $G_s$, as given by \eqref{eq:modal_temp} and \eqref{eq:frozen_z}, is obtained.

In this paper, the delays from the hold circuit in the digital measurement environment are first carefully determined using a combination of readily available automatic methods and visual inspection of the data while manually tweaking the delay value. Subsequently the the FRF measurements are compensated for these delays such that the obtained delay-compensated FRF can be modeled in continuous time, i.e., by $\tilde{G}_s(s_k)$, with $s_k = j\omega_k$, and the remaining identification procedure can be performed in the $s$-domain. For additional details on such a pseudo continuous time modeling approach, see, e.g., \cite{Houpis1992}, \cite[Section 8.5]{PintelonSchoukens12}. Note that when the model is to be used in a control context, the importance of determining the correct delay increases significantly and one might consider changing to a fully discrete-time $z$-domain approach which is generally better suited to deal with such unknown delays.

\subsubsection{Modal Parametrization}\label{Sect:Modal_param}
As outlined in Section \ref{Sect:ID_approach}, the modal model for the spatially sampled system $G_s$ is given by \eqref{eq:modal_temp} and \eqref{eq:frozen_z}, with parameters contained in the matrices ${\mathcal{L}}_s,\,\Omega^2,\,\mathcal{D}_m,$ and $\mathcal{R}$. The matrices ${\mathcal{L}}_s$ and $\mathcal{R}$ are fully parameterized while $\Omega^2 = \mathrm{diag}(\bar{\omega}^2)$ with parameters $\bar{\omega}^2 = [\omega_1^2 \;\ldots \; \omega_{n_m}^2]\tran$. The damping matrix $\mathcal{D}_m$ is either fully parameterized, which constitutes a general viscous damping model, or, when using the modal damping model, this is equal to $\mathcal{D}_{m,\mathrm{mod}} = \mathrm{diag}(\bar{\zeta})$ with $\bar{\zeta} = [\zeta_1 \;\ldots \; \zeta_{n_m}]\tran$.

Using the modal damping model, the set of differential equations \eqref{eq:modal_temp} describing the systems temporal behavior becomes fully decoupled, meaning the system can be considered as a superposition of independently evolving modes \cite{Gawronski1998}. This representation is extensively used in modal analysis and design as it simplifies the physical interpretation of the modal parameters and the incurred modeling errors by assuming modal damping is generally small for lightly damped systems, see \cite[Section 2.4]{Gawronski1998}. For the wafer-stage example the modal damping assumption is used to facilitate a parsimonious parametrization. Note that the modeling framework used in this paper enables general linear damping models.

The complexity of this parametrization is determined by the number of modes that are modeled, $n_m$. While a limited number of modes usually dominate the behavior in a given frequency range of interest, for some applications the combined low-frequency compliance effect of unmodeled higher order modes also needs to be taken into account. In such a case it is relevant to model an additional compliance term, e.g., as direct feed-through such as $D(\varrho)$ in \eqref{eq:spa_temp_2}, to describe the quasi-static deformation resulting from each input signal $u(t)$, see, e.g., \cite{Voorhoeve2016a}. In this paper such a feed-through term is not modeled; this can be straightforwardly incorporated in the proposed modeling framework when required for a given application.

The modal parametrization used in this paper is defined by \eqref{eq:modal_temp} and \eqref{eq:frozen_z} with the modal parameters given by,
\begin{equation} \label{eq:modal_parameters}
\theta_m = \mathrm{vec}\left(\begin{bmatrix}
\mathcal{L}_s\tran & \mathcal{R} & \bar{\omega}^2 & \bar{\zeta}
\end{bmatrix}\right)\,.
\end{equation}

\subsubsection{MFD Parametrization}\label{Sect:MFD_param}

For certain identification algorithms, it is necessary that the model parametrization can be written as a polynomial matrix fraction description. In this paper, a Left Matrix Fraction Description (LMFD) parametrization is used, given by

\begin{equation}
\label{eq:MFD_CH6}
\hat{G}(s,\theta) = \hat{D}(s,\theta)^{-1}\,\hat{N}(s,\theta)\, ,
\end{equation}
where $\hat{N}(s,\theta) \in \mathbb{R}^{p\times q}[s]$ and  $\hat{D}(s,\theta) \in \mathbb{R}^{p\times p}[s]$ are real polynomial matrices in the Laplace variable $s$. Furthermore, these polynomial matrices are linearly parameterized with respect to the parameter vector $\theta \in \mathbb{R}^{n_\theta}$ using a set of basis functions such that,
\begin{align}
\label{eq:basis_CH6}
&\mathrm{vec}\left( \begin{bmatrix} \hat{D}(s,\theta) & \hat{N}(s,\theta) \end{bmatrix} \right) = \sum_{j=1}^{n_\theta} \psi_j(s) \theta_j = \Psi(s) \theta\, .
\end{align}
This general linear parametrization allows the use of data-dependent orthogonal vector polynomials as basis functions, $\psi_j \in \mathbb{R}^{p(q+p) \times 1}[s]$, see, e.g., \cite{Herpen2014aut}, which is a key aspect for the identification of increasingly complex system where the numerical conditioning becomes an important limiting factor for the performance of the identification algorithms.

Furthermore, this general parametrization enables the use of more structured LMFD parameterizations, which enable the parametrization of system with arbitrary McMillan degree $n_x$ instead of only being able to parameterize systems where the degree $n_x$ is a multiple of the number of outputs $p$, as is the case when using the fully parameterized unstructured LMFD as in, e.g., \cite{guillaume2003,Cauberghe2004}. Here, a generic (pseudo)-observable-canonical LMFD parametrization with a {quasi-constant} degree structure is used, see, e.g., \cite{GloverWillems74,Vayssettes2014,Mathelin1991}. This parametrization is both identifiable, in the sense that it is not over-parameterized, and generic, meaning that it can approximate all proper LTI systems of the given order up to arbitrary precision, see \cite{GloverWillems74}.

This LMFD parametrization is often used for black-box identification of LTI systems, whereas in this paper the goal is to identify spatio-temporal mechanical systems by utilizing the modal form, i.e., \eqref{eq:modal_temp} and \eqref{eq:frozen_z}. Therefore, in this paper a number of additional constraints are incorporated in the LMFD parametrization such that it more closely resembles the mechanical system model. Here, the following properties are enforced,
\begin{enumerate}
\item an even McMillan degree, by taking $n_x = 2\, n_m$,
\item a relative degree $r \geq 2$, by constraining the column degrees of the numerator polynomial matrix $N(s,\theta)$ to be $2$ lower than the corresponding column degrees of the denominator polynomial matrix $D(s,\theta)$,
\item a prescribed number of rigid-body modes $n_\mathrm{rb}$ such that $n_0 = 2\, n_\mathrm{rb}$ poles are located at $s=0$, by factoring out the rigid body dynamics, see \cite{Voorhoeve2016} for details.
\end{enumerate}

\subsubsection{Identification Algorithms}
The identification problem considered here is to find the parameter vector $\theta$ that minimizes the identification cost, which is in this paper is a weighted least-squares cost function, i.e.,
\begin{equation}
\label{eq:crit_CH6}
\hat{\theta} = \argmin\limits_\theta\, V(\theta) = \sum_{k=1}^m {\varepsilon}(s_k,\theta)^H {\varepsilon}(s_k,\theta)\, ,
\end{equation}
where
\begin{equation} \label{eq:crit_2}
{\varepsilon}(s_k,\theta) = W(k)\,\,\mathrm{vec}\left({\tilde{G}}(s_k)-\hat{G}(s_k,\theta) \right) \, ,
\end{equation}
with weighting matrix $W(k) \in \mathbb{C}^{pq \times pq}$.
This general cost function $V(\theta)$ includes other commonly used identification criteria \cite{Voorhoeve2016}, such as the sample maximum likelihood criterion \cite{PintelonSchoukens12}, control relevant identification criteria \cite{Oomen2014}, and the input-output and element-wise weighted criterion used in \cite{CallafonRooHof1996}.

Minimization of the cost function \eqref{eq:crit_CH6} is a nonlinear least-squares optimization problem. Suitable algorithms to solve this problem are the Sanathanan-Koerner algorithm and the Gauss-Newton algorithm, or closely related methods such as the Levenberg-Marquart algorithm. These algorithms are defined as follows, where the Sanathanan-Koerner algorithm is only defined for MFD parameterizations, i.e., using \eqref{eq:MFD_CH6}--\eqref{eq:basis_CH6}.

\begin{myalg}[Sanathanan-Koerner \cite{SanathananKoe1963}]
Let $\theta^{\langle 0 \rangle}$ be given. In iteration $i = 0, 1, \dots$, solve the linear least squares problem
\begin{equation}
\theta^{\langle i+1\rangle} = \argmin\limits_{\theta} \sum_{k = 1}^m \left\Vert  W_\mathrm{SK}(s_k, \theta^{\langle i\rangle})\, \Psi(s_k)\theta \right\Vert_2^2 ,
\label{eq:Alg_SK_OAT}
\end{equation}
with
\begin{equation}
W_\mathrm{sk}(k, \theta^{\langle i\rangle}) = W(k) \left(\begin{bmatrix} \tilde{G}\tran(s_k) & -I_q \end{bmatrix} \otimes \hat{D}(s_k,\theta^{\langle i\rangle})^{-1}\right).
\end{equation}
\end{myalg}

\begin{myalg}[Gauss-Newton \cite{Bayard1994}] \label{alg:GN_CH6}
Given an initial estimate $\theta^{\langle 0 \rangle}$, compute for $i = 0, 1, \dots$
{ \begin{equation}
\theta^{\langle i+1\rangle} = \theta^{\langle i \rangle} + \argmin_{\Delta_\theta} \sum_{k = 1}^m \left\Vert J(s_k,\theta^{\langle i \rangle})\Delta_\theta + \varepsilon(s_k,\theta^{\langle i \rangle}) \right\Vert_2^2,
\label{eq:Alg_GN_CH6}
\end{equation}}
with
\begin{equation}
J(s_k,\theta^{\langle i \rangle}) = \left.\dfrac{\partial \varepsilon(s_k,\theta)}{\partial \theta\tran}\right|_{\theta^{\langle i \rangle}} = -W(k)\left.\dfrac{\partial\, \mathrm{vec}({\hat{G}}(s_k,\theta))}{\partial\theta\tran}\right|_{\theta^{\langle i \rangle}}\!.
\end{equation}
\end{myalg}

The Gauss-Newton algorithm, and related algorithms such as the Levenberg-Marquardt algorithm, generally provide fast monotonic convergence to a minimum of the cost function $V(\theta)$. However, these algorithms often converge to local-minima that are far from optimal, and therefore their performance is strongly dependent on the quality of the initial estimate $\theta^{\langle 0 \rangle}$. The Sanathanan-Koerner algorithm on the other hand does not generally converge monotonically, and, if convergent, its stationary points are generally not optima of the cost function \cite{Whitfield87}. However, the Sanathanan-Koerner algorithm often yields adequate, albeit suboptimal, results irrespective of the quality of the initial estimate. Therefore, the Sanathanan-Koerner algorithm is often used to provide initial estimates that are subsequently refined using a gradient based optimization algorithm such as the Gauss-Newton algorithm, see, e.g., \cite{Voorhoeve2016}.

\subsubsection{Identification Approach}
\label{Sect:LTI_ID_approach}
To identify the modal system model defined by \eqref{eq:modal_temp}, \eqref{eq:frozen_z} and parameterized by \eqref{eq:modal_parameters} from the identified FRF $\tilde{G}_s(s_k)$, the following steps are followed.
\begin{enumerate}
\item Define weighting filters $W(k)$ as in \eqref{eq:crit_2}.
\item Perform $i_\mathrm{SK}$ iterations of the SK algorithm using the mechanical LMFD model with constraints as defined in Section \ref{Sect:MFD_param}.
\item Perform $i_\mathrm{GN}$ iterations of the GN or LM algorithm for the LMFD model using the parameters $\theta_\mathrm{SK,min}$ corresponding to the lowest cost function value during SK iterations as initial estimate.
\item Transform the identified LMFD model to an initial estimate for the modal model as defined by parameters \eqref{eq:modal_parameters}.
\item Perform a maximum of $i_\mathrm{GN,mod}$ iterations of the GN or LM algorithm to converge to a optimum of the cost function as in \eqref{eq:crit_CH6} for the identified modal model.
\end{enumerate}

When considering generally damped modal models, the fourth step of this identification approach can be performed using an exact transformation, i.e., relating two realizations of the same system. Details of this exact transformation are beyond the scope of this paper and will be reported elsewhere. Due to the modal damping assumption used in this paper for the modal model, and since this modal damping assumption is not enforced in the MFD parametrization, the transformation in step 4 of this identification approach is approximate. This approximate transformation is performed by calculating the $n_x = 2 n_m$ pole locations by solving $\det[D(s,\theta)] = 0$, separating the poles into pole pairs such that $(s+p_{1,i})(s+p_{2,i}) = s^2 + \zeta_i s+ \omega_i^2$, with $\zeta_i,\,\omega_i \in \mathbb{R}$ for $i=1,\ldots,\,n_m$, and estimating a model of the form,
\begin{equation}
\hat{G}_{m,\mathrm{trans}} = \sum_{i=1}^{n_m} \dfrac{R_i}{s^2 + \zeta_i s+ \omega_i^2}\,,
\end{equation}
with $R_i \in \mathbb{R}^{p\times q}$.  In this estimation the denominator parameters are fixed to the values obtained from the pole-pairs of the LMFD model. This model is fitted based on the FRF data using the same cost function as the other identification steps, i.e, using \eqref{eq:crit_CH6}. The parameters in $\mathcal{L}$ and $\mathcal{R}$ are then obtained from the singular value decomposition of the residue matrices $R_i = U_iS_iV_i^H$ such that
\begin{align}
[\mathcal{L}_s]^i &= [U_i]^1[S_i]_1^1, \quad [\mathcal{R}]_i = [V_i^H]_1, \quad i = 1, \dots, n_m, \label{eq: D SVD R}
\end{align}
where $[X]^i$ and $[X]_j$ denote, respectively, the $i$-th and column and the $j$-th row of matrix $X$. This transformation performs well for the considered system, as shown by the results in Section \ref{Sect:Results_ID}. For more general approaches to transform black-box models to a gray-box models, see, e.g., \cite{Mercere2014}.

\begin{figure}[t]
\begin{center}
\includegraphics[width=1\linewidth]{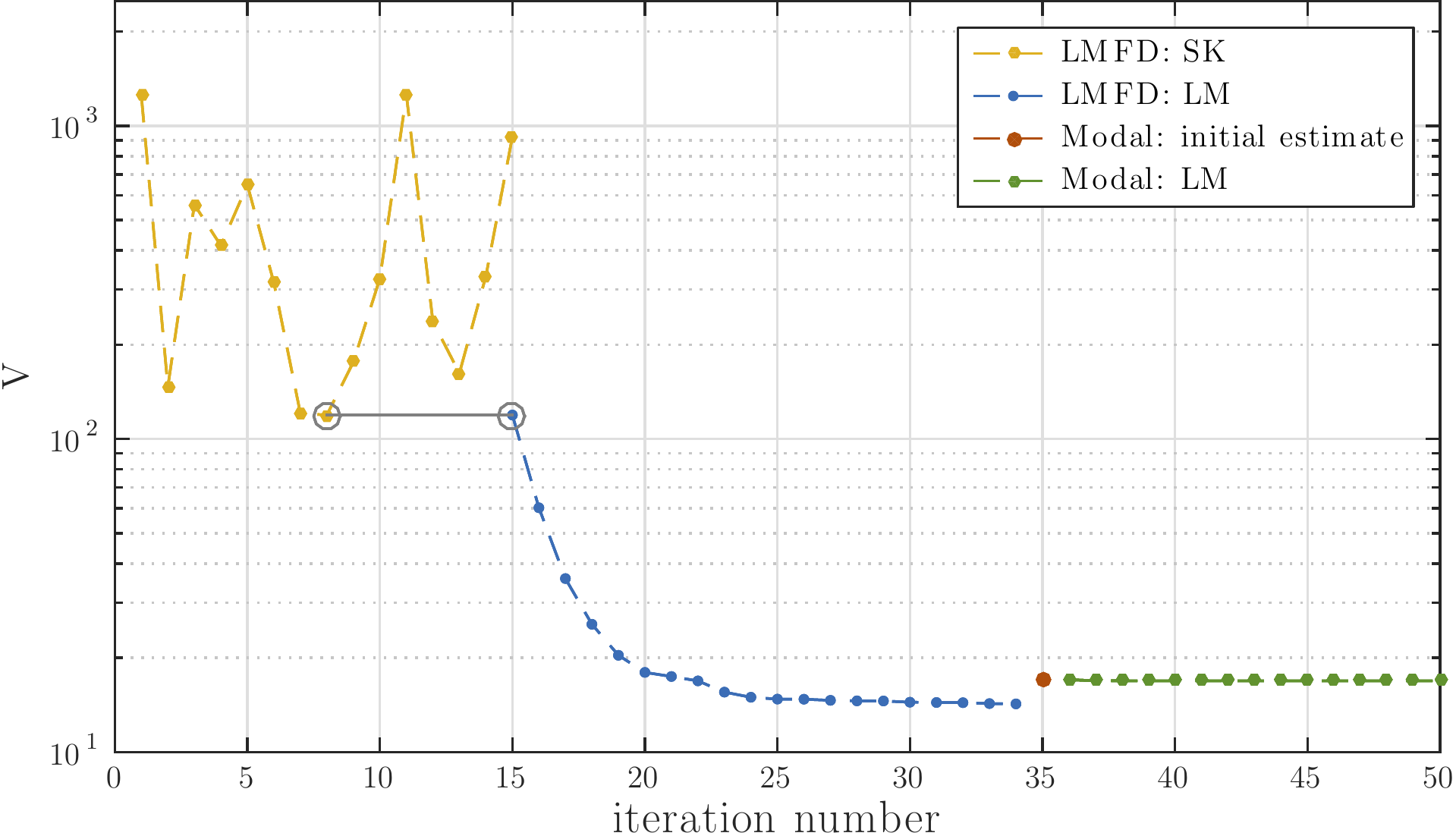}
\caption{Evolution of the cost function during various steps of the LTI identification approach.}
\label{fig:conv}
\end{center}
\end{figure}

\subsection{Results}
\label{Sect:Results_ID}
\begin{figure*}
\begin{center}
\includegraphics[width=1\linewidth]{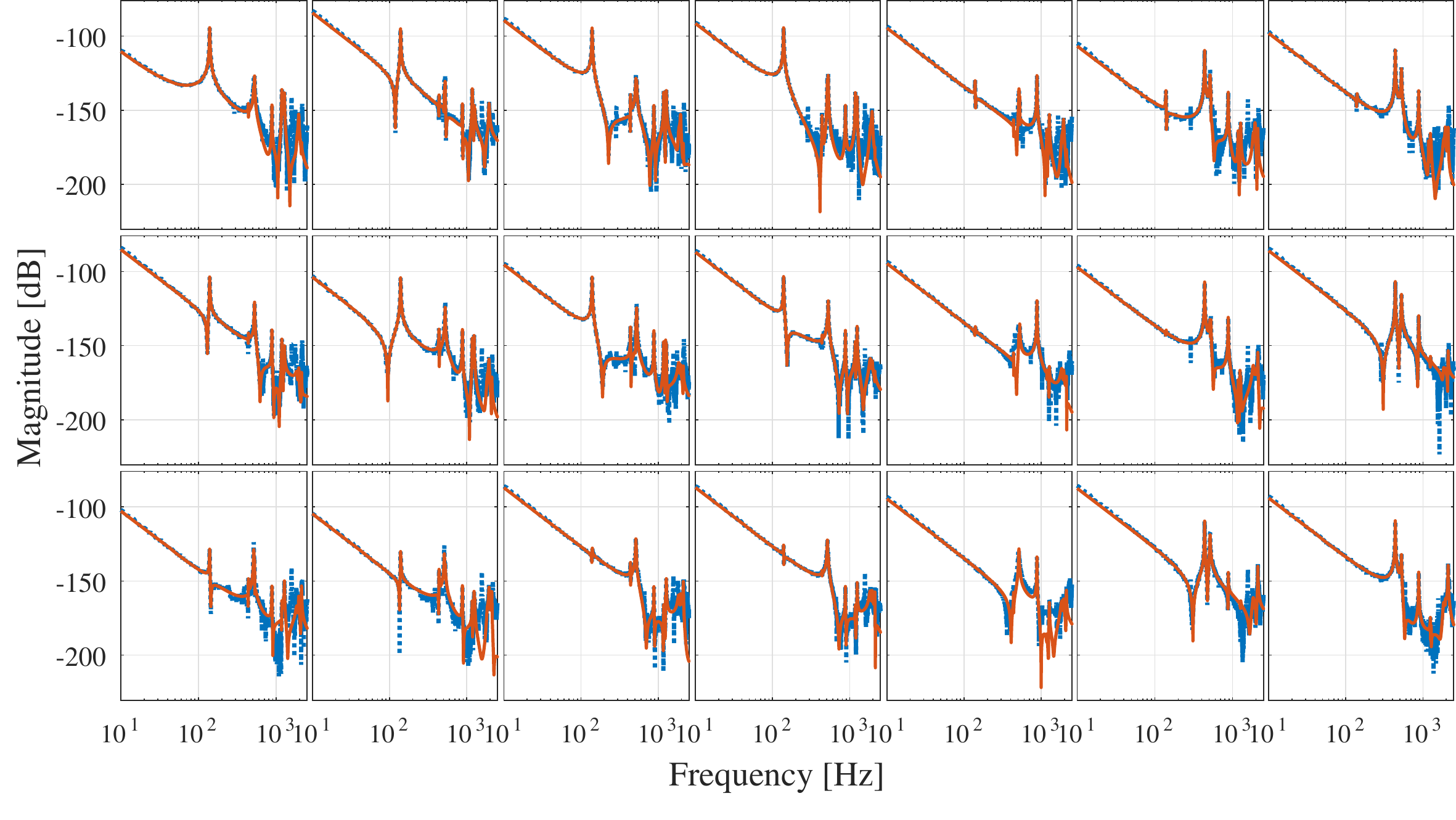}
\caption{Bode diagrams of the identified FRFs (dotted blue) and fitted modal model (solid red) for a $3\times 7$ part of the full $16 \times 7$ identified system.}
\label{fig:FRFs_and_fit}
\end{center}
\end{figure*}

For the identification of the wafer-stage system, the weighting function in \eqref{eq:crit_2} is chosen as a weighting with the element-wise inverse of the identified FRF $\tilde{G}(s_k)$, i.e.,
\begin{align}
W_\mathrm{inv}(k)&=\mathrm{diag}(\mathrm{vec}(\left|\Lambda(s_k)\right|))\,,\\
[\Lambda(s_k)]_j^i &= \dfrac{1}{[\tilde{G}(s_k)]_j^i}\,.
\end{align}
This choice reflects the goal of minimizing the relative error between the model $\hat{G}(s_k,\theta)$ and the FRF $\tilde{G}(s_k)$. For more advanced weighting choices that take into account the control objective, see \cite{Oomen2014}. To emphasize the accurate estimation of the first few lower-frequency resonances the weighting function is truncated, i.e.,
\begin{equation}
W(k) = \min\left(W_\mathrm{inv}(k),\,w_\mathrm{max}\right)\,,
\end{equation}
where $w_\mathrm{max}$ is chosen such that clipping of the weight generally occurs only in the high frequency range, after the first few resonances.

Steps 2-5 of the identification approach as proposed in Section \ref{Sect:LTI_ID_approach} are at first only performed  for a $3 \times 7$ part of the full $16 \times 7$ identified FRF. This is done both to improve computational efficiency and to simplify the implementation of rigid-body mode constraints, see, e.g., \cite{Voorhoeve2016}. After the successful identification of the modal parameters for the $3 \times 7$ system, the model for the remaining $13$ outputs is determined by estimating the sampled mode-shape parameters in $\mathcal{L}_s$ for these outputs while all other parameters remain fixed. This is again done by minimizing the cost function \eqref{eq:crit_CH6} for these additional output, which in this case is simply a linear least squares problem.

In Figure \ref{fig:conv}, the evolution of the cost function, $V(\theta)$ in \eqref{eq:crit_CH6}, is shown during steps 2-5 of the identification approach of Section \ref{Sect:LTI_ID_approach}. As can be seen from this figure, the SK algorithm in step 2 is not monotonically convergent, but yields an appropriate initial estimate for subsequent refinement using the LM algorithm. The LM algorithms used in step 3 does show monotonic convergence and yields an LMFD estimate with a cost function value approximately one order of magnitude below that of the initial estimate provided by the SK approach. In the next step the LMFD model is transformed to the modally damped model defined by the parameters \eqref{eq:modal_parameters}, which leads to a slight increase in the cost function value. In the final step the cost function is again minimized using the monotonically convergent LM algorithm with the modal parametrization, which in this case only yields a small decrease of the cost function value showing that the initial modal estimate is already of a high quality.

The small increase in cost function value when transforming the LMFD modal to the modal model is expected, as this step reduces the model complexity by enforcing modal damping and through the elimination of computational modes identified in the LMFD model. This is done by visually comparing the identified pole-locations with the resonances of the identified FRFs, similar to the use of stabilization diagrams in experimental modal analysis, see, e.g., \cite{guillaume2003,Cauberghe2004}. The model order of the identified LMFD model is $n_x = 2 n_m = 46$ while the model order of the modal model is $n_x = 24$. This shows that the transformation yields a significant decrease in model complexity with a modest increase in cost function value.

\begin{figure}
\begin{center}
\includegraphics[width=1\linewidth]{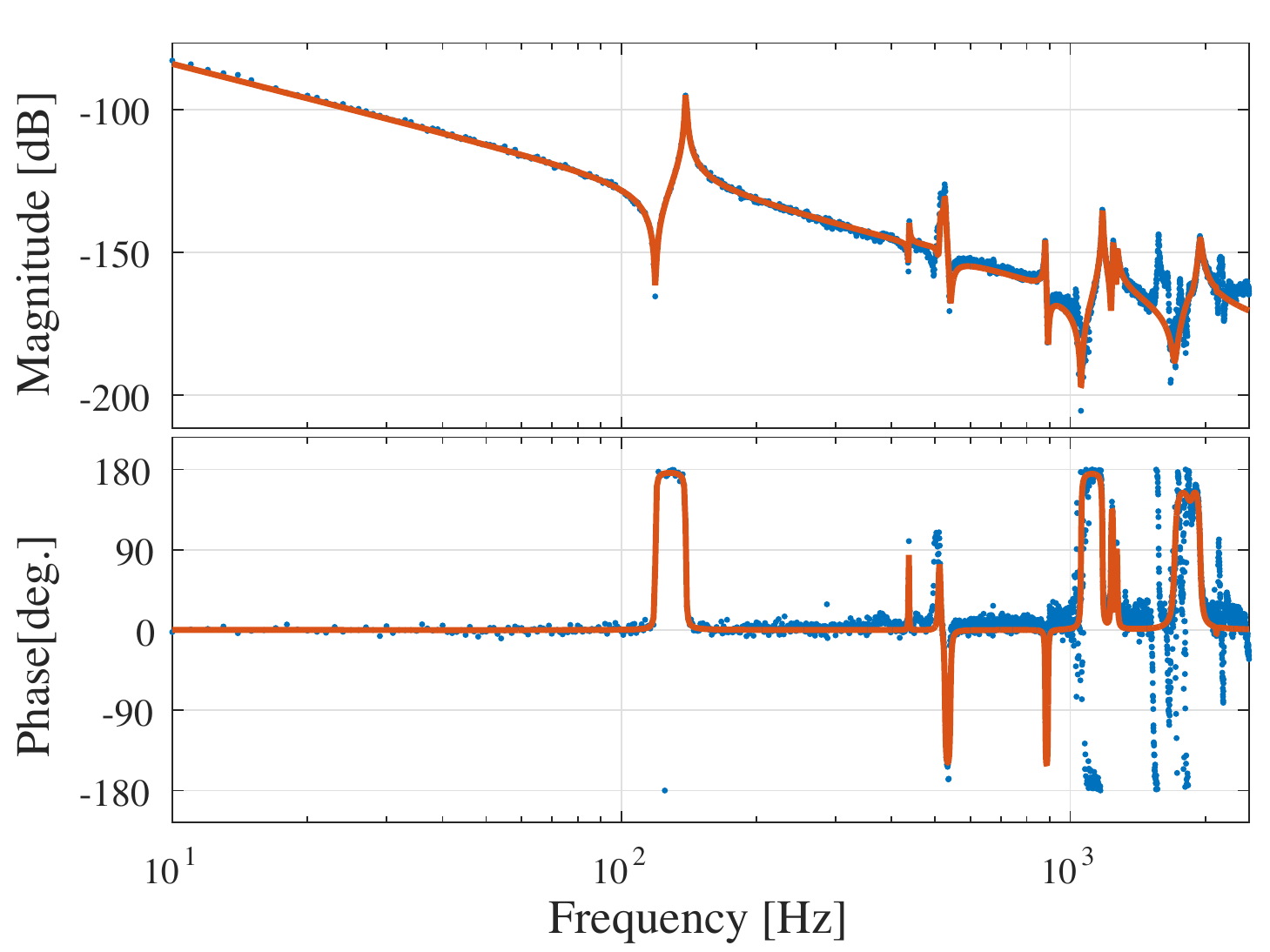}
\caption{Detailed Bode diagram including phase of the identified FRF (dotted blue) and fitted modal model (solid red) for the $(1,\,2)$ element of the identified system as depicted in Figure \ref{fig:FRFs_and_fit}.}
\label{fig:FRF_with_phase_and_fit}
\end{center}
\end{figure}

Figure \ref{fig:FRFs_and_fit} shows the FRF and identified modal model for the $3 \times 7$ part of the system on which the identification procedure is performed and Figure \ref{fig:FRF_with_phase_and_fit} shows a more detailed view including the phase of a single element of the FRF and identified modal model. These figures show a good agreement between the model and the FRF in the low frequency region as well as for the dominant resonances in the frequency region up to about $1$ kHz. In the frequency region beyond $1$ kHz, there are an additional few accurately identified modes but also a number of unmodeled resonances. The identification of these high frequency modes is not the main focus in this paper since the spatial behavior for such high frequency modes is typically also of a higher spatial frequency, meaning a higher spatial resolution, than what is available, is required to identify the associated mode-shapes.

\section{Mode-Shape Interpolation}
\label{Sect:interp_OAT}

In this section, the interpolation of the spatially sampled mode shapes, as given by the columns of $\mathcal{L}_s$, is considered. This interpolation step is performed to obtain a position-dependent model that is continuous in the spatial variable $\varrho$.

\subsection{Methods}

\begin{figure*}
\begin{subfigure}{.48\linewidth}
\includegraphics[width=1\linewidth]{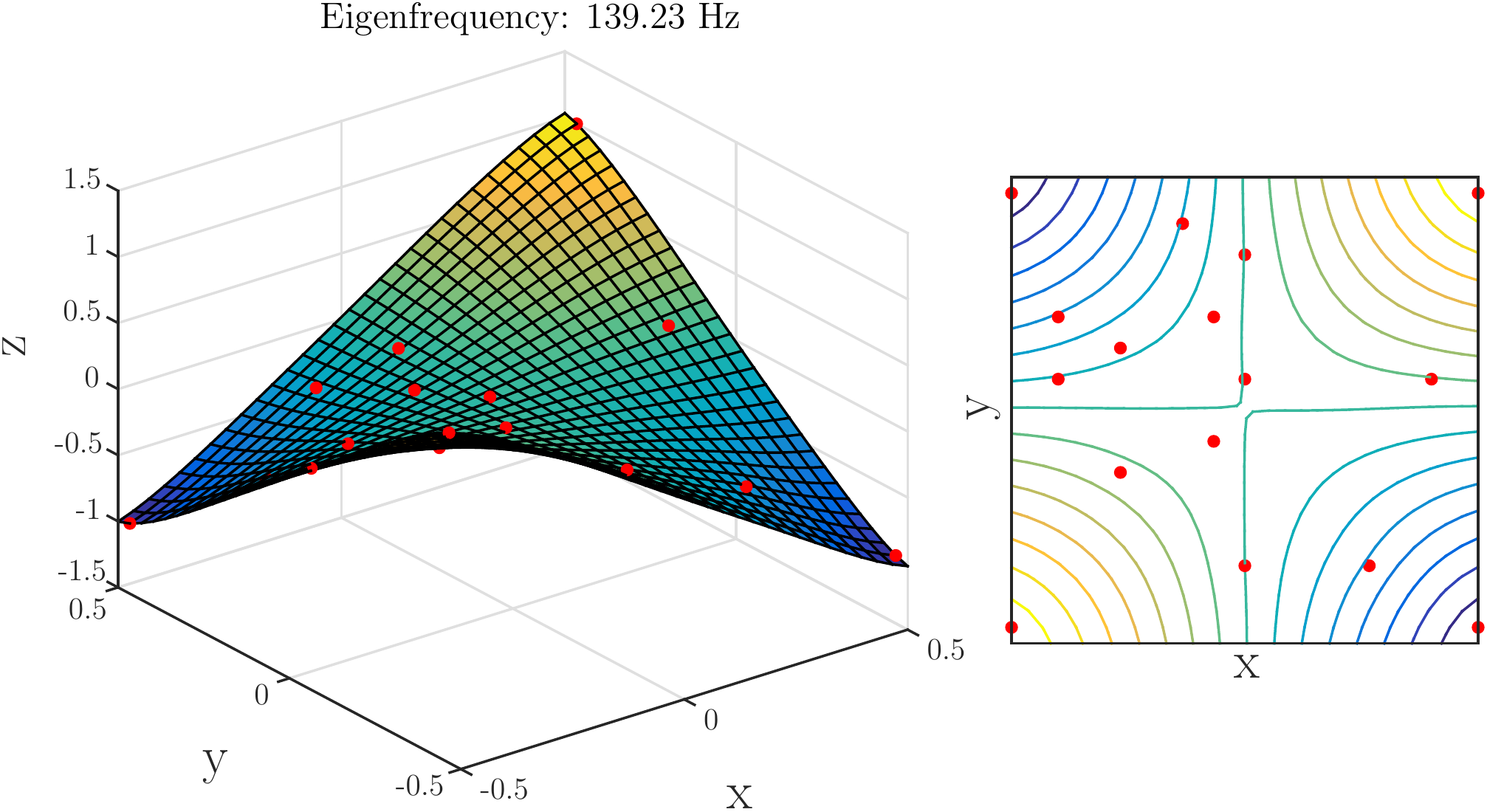}
\caption{Fourth mode (torsion).}
\label{fig:Mode4}
\end{subfigure}\hspace{.035\linewidth}
\begin{subfigure}{.48\linewidth}
\includegraphics[width=1\linewidth]{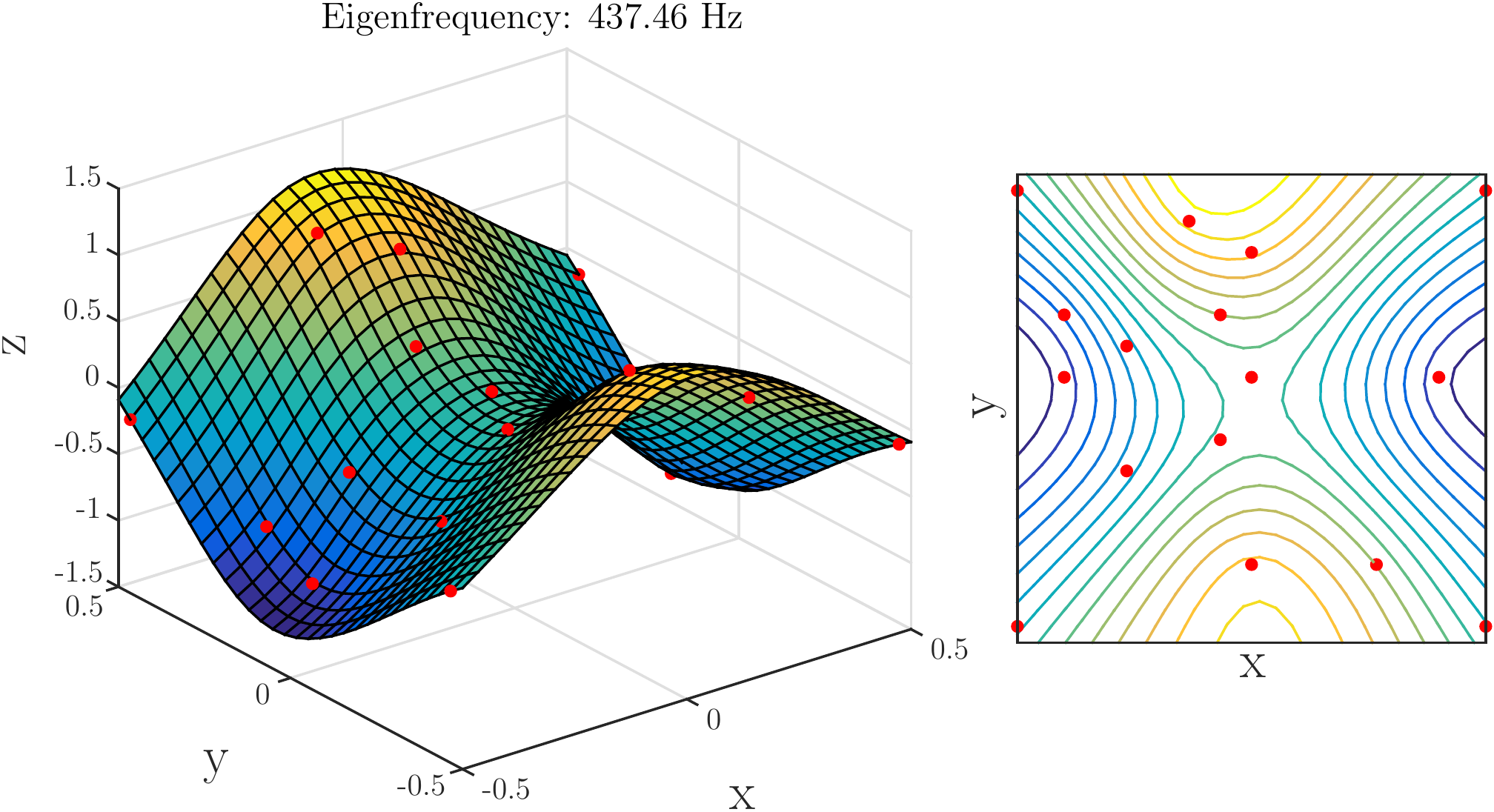}
\caption{Fifth mode (saddle).}
\label{fig:Mode5}
\end{subfigure}\\[1em]
\begin{subfigure}{.48\linewidth}
\includegraphics[width=1\linewidth]{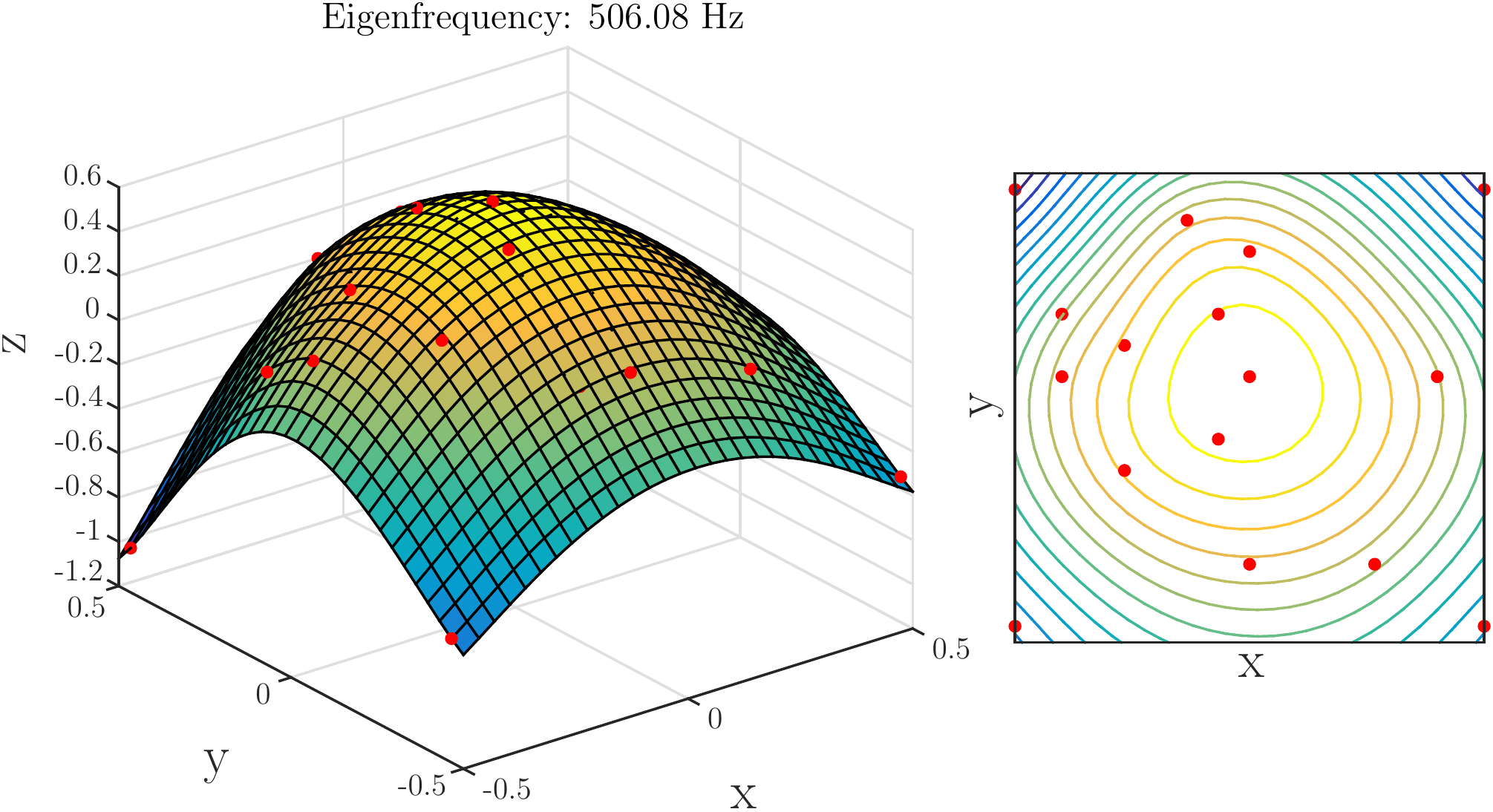}
\caption{Sixth mode (umbrella).}
\label{fig:Mode6}
\end{subfigure}\hspace{.035\linewidth}
\begin{subfigure}{.48\linewidth}
\includegraphics[width=1\linewidth]{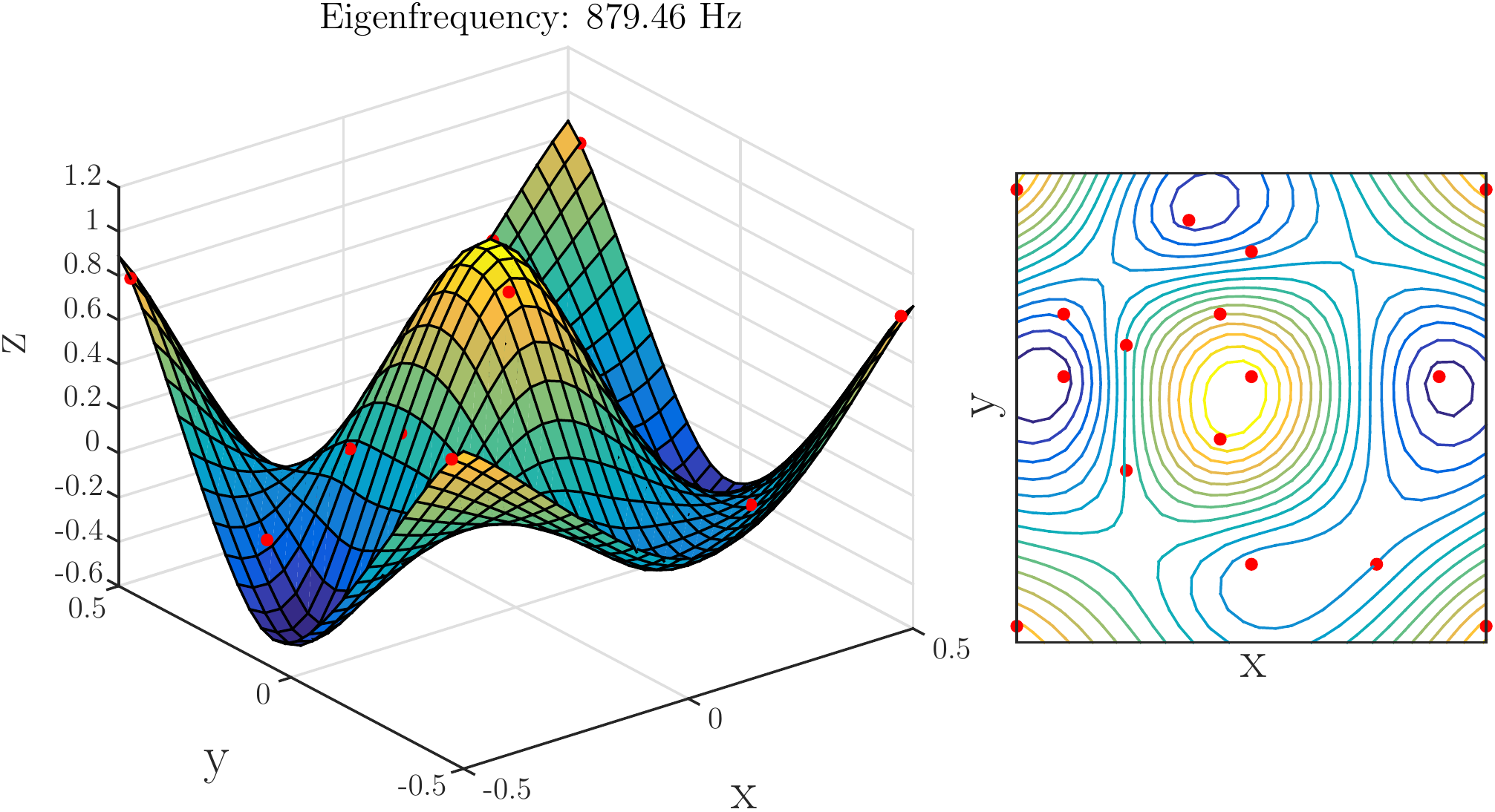}
\caption{Ninth mode.}
\label{fig:Mode9}
\end{subfigure}
\caption{Top views and 3-D surface plots of the identified mode shapes with red dots indicating the points of the identified spatially-sampled mode shapes.}
\label{fig:Modes}
\end{figure*}
%

A popular method for the interpolation of various types of data at arbitrary spatially distributed points is the smoothed thin-plate-spline interpolation approach. The use of thin-plate-splines is physically motivated by the fact that the spline functions are derived as the functions that minimize the bending energy of a thin sheet of elastic material. Therefore, this approach is particularly well-suited for the considered application of interpolating the structural mode-shapes of motion systems which are thin in one dimension relative to the other dimensions, such as the wafer-stage example.

The smoothed thin-plate-spline interpolating function $\mathcal{W}_s$ for a single mode-shape is derived as follows. Given a set of $n_\varrho$ points $\{(\bar{x}_j,\bar{y}_j,\bar{z}_j) \in \mathbb{R}^3\}$ and a user-defined smoothing parameter $\lambda \in [0,\infty)$, find an interpolating function $\mathcal{W}_s \in \mathfrak{W}_2^1$ such that,
\begin{align}
\min_{\substack{\mathcal{W}_s \in \mathfrak{W}_2^1}} \sum_{j = 1}^{n_{\varrho}} | \mathcal{W}_s(\bar{x}_j,\bar{y}_j) - \bar{z}_j |^2  +  \lambda U, \label{eq: STPS cost}
\end{align}
with
\begin{align}
U &= \int\limits_{-\infty}^{\infty} \int\limits_{-\infty}^{\infty} \Delta^2 \mathcal{W}_s(x,y)\, dx\, dy\,,
\end{align}
Here, the function space $\mathfrak{W}_2^1$ is the space of continuously differentiable functions with square-integrable second derivatives such that $U$, which is generally interpreted as a measure of the bending energy of the functions, exists for all functions in the space.

The functions $\mathcal{W}_s$ that minimize \eqref{eq: STPS cost} are given by
\begin{align}
&\mathcal{W}_s(x,y,\vartheta) = \vartheta_0 + x \vartheta_x + y \vartheta_y + \sum_{j = 1}^{n_{\varrho}} \vartheta_{j} G_j(x,y), \quad  \label{eq: TPS par}\\
&G_j(x,y) = r_j^2 \text{ln}(r_j), \quad r_j = \sqrt{(\bar{x}_j-x)^2+(\bar{y}_j -y)^2}\,, \label{eq: TPS sol}
\end{align}
see, e.g., \cite{Wahba1990}. The number of parameters in \eqref{eq: TPS par} is $n_\varrho+3$, where the three additional parameters are related to the monomials up to the first degree which represent the set of functions in $\mathfrak{W}_2^1$ for which $U=0$, i.e., the kernel of $U$. To constrain this underdetermined set of equations, the following three additional constraints are added which make sure that the function space parameterized using the Green's functions $G_j(x,y)$ is orthogonal to the space of first order polynomial,
\begin{equation}
\sum_{j=1}^{n_\varrho} \vartheta_j = 0\,,\quad \sum_{j=1}^{n_\varrho} \vartheta_j\bar{x}_j = 0\,,\quad \sum_{j=1}^{n_\varrho} \vartheta_j\bar{y}_j = 0\,. \label{eq: TPS constr}
\end{equation}
The solution to \eqref{eq: STPS cost} using \eqref{eq: TPS par} and \eqref{eq: TPS constr} is given by
\begin{equation}
\bar{z}_k = \mathcal{W}_s(\bar{x}_k,\bar{y}_k,\vartheta) + \lambda \vartheta_k\,,
\end{equation}
see, e.g., \cite{Wahba1990}. In explicit matrix-form this yields
\begin{align}
\vartheta &= X^{-1}\begin{bmatrix}
\bar{z}_1 & \dots & \bar{z}_{n_\varrho} & \mathbf{0}_{1\times 3}
\end{bmatrix}\tran\,,
\end{align}
with $\vartheta = \begin{bmatrix} \vartheta_0 & \vartheta_x & \vartheta_y & \vartheta_1 \; \dots\; \vartheta_{n_\varrho} \end{bmatrix}\tran$, and where
\begin{align}
X &= \begin{bmatrix}
X_0 & X_G + \lambda I \\ \mathbf{0}_{3 \times 3} & X_0\tran
\end{bmatrix},\quad
X_0 = \begin{bmatrix}
1 & \bar{x}_1 & \bar{y}_1 \\
 & \vdots &\\
1 & \bar{x}_{n_\varrho} & \bar{y}_{n_\varrho}
\end{bmatrix},\\
X_G &= \begin{bmatrix}
G_1(\bar{x}_1,\bar{y}_1) & \dots & G_{n_\varrho}(\bar{x}_1,\bar{y}_1)\\
\vdots & & \vdots\\
G_1(\bar{x}_{n_\varrho},\bar{y}_{n_\varrho}) & \dots & G_{n_\varrho}(\bar{x}_{n_\varrho},\bar{y}_{n_\varrho})\\
\end{bmatrix}\,.
\end{align}

For the interpolation of the spatially-sampled mode shapes as identified in $\mathcal{L}_s$, the points $(\bar{x}_j,\bar{y}_j)$ are equal to  $\varrho_j$, i.e., the $(x,y)$ positions of the sensors, and the values for $\bar{z}_j$ are given by the identified parameters in the columns of $\mathcal{L}_s$. This interpolation is carried out independently for each mode shape, i.e., for each column of $\mathcal{L}_s$, where for each mode shape a different smoothing parameters $\lambda$ is used. These smoothing parameters provide a trade-off between robustness to estimation errors in $\mathcal{L}_s$ and interpolation accuracy at the data-points $\varrho_j$. In this paper, the values of the smoothing parameters are determined using a Leave-One-Out-Cross-Validation (LOOCV) approach, i.e., the value of $\lambda$ is used which minimizes the LOOCV error. For details on this cross-validation approach, see, e.g., \cite{Wahba1980}.

\subsection{Results}
In Figure \ref{fig:Modes}, four of the identified flexible mode-shapes are shown. In total $12$ mode-shapes are identified including the three out-of-plane rigid-body modes. Up to the ninth mode, as shown in Figure \ref{fig:Mode9}, the identified mode shapes agree well with theoretical mode-shapes for a thin-plate or the mode-shapes as obtained by means of a Finite-Element-Method analysis of the system, a detailed comparison is omitted for brevity. For the higher-order modes the spatial resolution of the sensors is insufficient to accurately reconstruct the smooth mode-shapes.

The results in Figure \ref{fig:Modes} show the viability of the proposed approach to obtain accurate position-dependent models of flexible mechanical systems. In the following section, the potential of the proposed approach is discussed in enabling various position-dependent control approaches.

\section{Outlook for Position-Dependent Motion Control applications}
\label{Sect:Control_app}

In this section, several control approaches are explored that are enabled by the availability of accurate position-dependent models. For flexible motion systems both feedback and feedforward control problems become more complex as the point that should track the reference is often not directly measured, such as the point-of-interest of the wafer stage example in Figure \ref{fig:wafer_schem}. Furthermore, the location of this point-of-interest can change over time, increasing the complexity of the control problem.

Approaches to enhance the control performance for such flexible motion systems can generally be classified as either,
\begin{enumerate}
\item global approaches, aimed at preventing or mitigating the flexible deformations in the entire system, such that the deformation related errors are small; or
\item local approaches, aimed at controlling the position of the point-of-interest of the deformed system
\end{enumerate}
By utilizing the identified position-dependent model of the wafer-stage system, $\hat{G}_m(\varrho)$, such global and local control approaches can be described in the LPV standard plant framework, as shown in Figure \ref{fig:wafer_contr}, and can be solved using a range of approaches. In this section, several global and local approaches are considered for both feedback and feedforward control.

\subsection{Global Spatio-Temporal Feedforward Control}
For the wafer-stage example, the problem of global feedforward control aims to minimize the error between the reference signal $r_z(t)$ and the out-of-plane deflection of the surface of the wafer stage $\mathfrak{z}(\varrho,t)$ over the entire spatial domain $\mathfrak{D}$. More precisely, the global approach is aimed at minimizing the following weighted spatial-norm of the error $e(\varrho,t) = r_z(t)-\mathfrak{z}(\varrho,t)$
\begin{equation}
\Vert e \Vert_{2(\mathfrak{D}_{W_S})} = \sqrt{\int\limits_{-\infty}^\infty \int_\mathfrak{D} e\tran(\varrho,t) W_S(\varrho)  e(\varrho,t)\, d\varrho\, dt}\,.
\end{equation}
This problem can be effectively formulated and solved as an $\mathcal{H}_\infty$ optimal control problem, for details see \cite{Rozario2017}.

\begin{figure}[t]
\begin{center}
\includegraphics[width=1\linewidth]{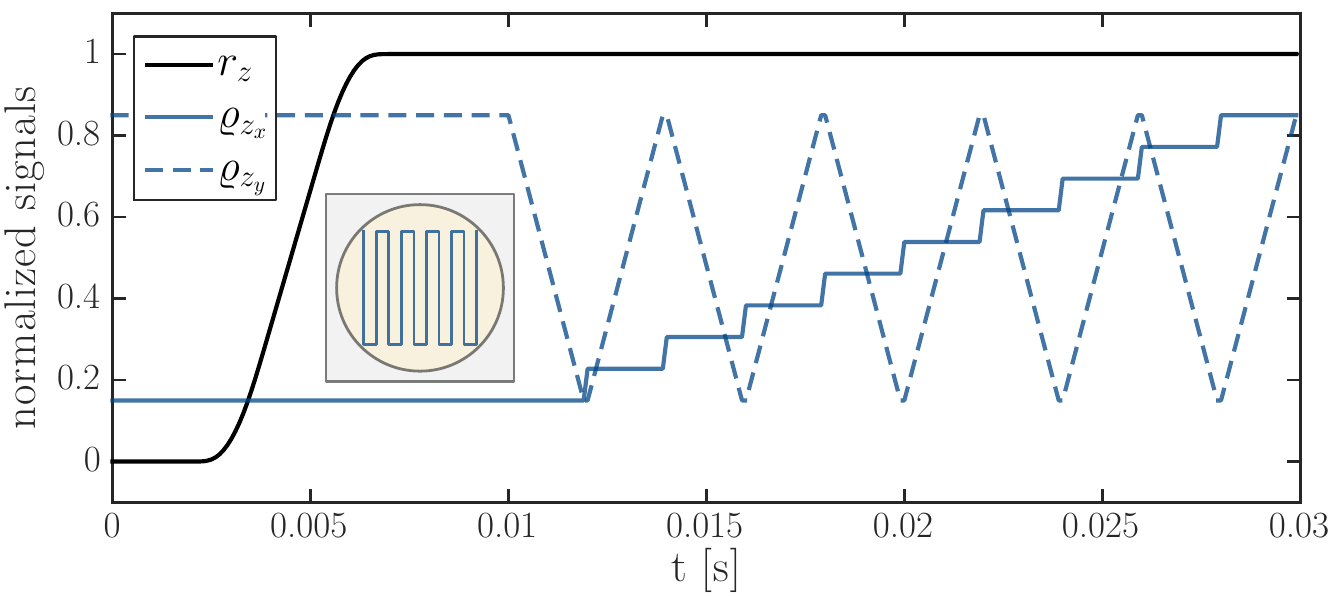}
\caption{Reference profile $r_z$, and the $(x,\,y)$-coordinates of the point of
interest $\varrho$ over time.\vspace{-1.5cm}}
\label{fig:ff_ref}
\end{center}
\end{figure}

\begin{figure}[t]
\begin{center}
\includegraphics[width=1\linewidth]{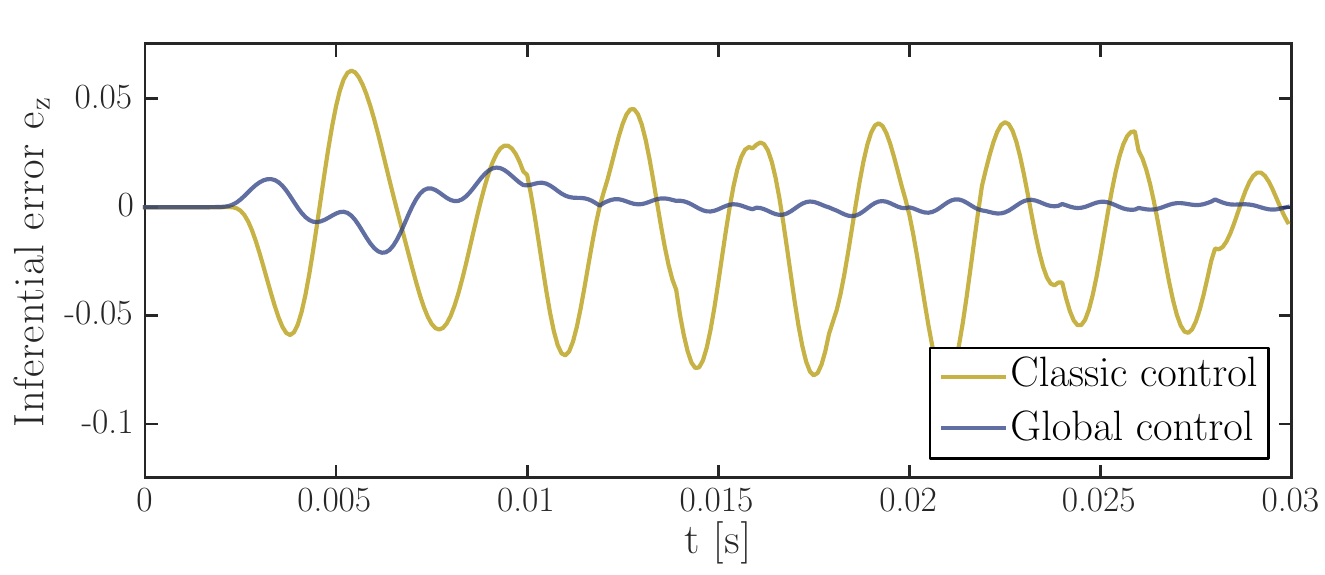}
\caption{Inferential error $e_{z}$ at point of interest over time, showing significantly improved performance for the global feedforward controller relative to the classical feedforward controller.\vspace{-.5cm}}
\label{fig:ff_global}
\end{center}
\end{figure}

In Figures \ref{fig:ff_ref} and \ref{fig:ff_global}, simulation results are shown for the wafer stage example system. Here, Figure \ref{fig:ff_ref} shows the reference profile for $r_z$ as well as the $(x,\,y)$ coordinates of the point of interest $\varrho_{z}(t)$ and Figure \ref{fig:ff_global} shows the local errors at the point of interest, i.e., $e_{z}(t) = r_z(t) - \mathfrak{z}(\varrho_{z}(t),t)$, for a classical feedforward controller, that minimizes the error at the sensor locations, and the global feedforward approach. These results show that the global approach yields a significantly improved inferential performance as opposed to the classical approach. These results and the practical potential of this global feedforward approach for future motion systems are enabled by the modeling procedure developed in this paper.

\subsection{Local Inferential Feedback and Feedforward Control}
\label{Sect:Control_fb}

The general problem of local inferential control aims to directly optimize the performance at the location where performance is required, such as the point-of-interest in the wafer-stage example, see Figure \ref{fig:wafer_schem}. Both for feedback and feedforward, these approaches can be described in the LPV standard plant framework when an accurately identified position-dependent model is available. Local inferential feedforward control typically involves the  inversion of the system dynamics from the inputs to the time-varying or parameter-varying performance variables which, apart from explicit inversion, can be realized by solving an optimal control problem or using iterative learning control, see, e.g., \cite{Hoffmann2015,Rozario2017a}. 
Inferential feedback control, generally requires the use of two degree of freedom controller structures as opposed to the single degree of freedom controller structure used in traditional feedback control, see, e.g., \cite{Oomen2015,Voorhoeve2016a}, this can be directly incorporated in the standard-plant approach.

\begin{figure}[t]
\begin{center}
\includegraphics[width=.65\linewidth]{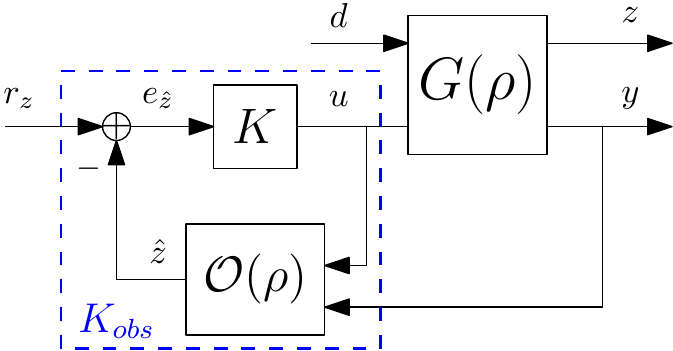}
\caption{Position-dependent observer-based feedback control scheme.}
\label{fig:fb_obs}
\end{center}
\end{figure}

Inferential feedback control is especially relevant when significant disturbance forces are present in the system. In \cite{Voorhoeve2016a}, it is shown that a disturbance-observer can be effectively utilized to estimate the inferential performance variable in the presence of significant disturbance forces that are non-collocated with the actuator forces. In  \cite{Voorhoeve2016a}, it is also shown that in such a case it is essential to include disturbance models in the standard-plant description to obtain accurate results. In \cite{Voorhoeve2016a}, an observer-based inferential control approach is proposed which is especially suited to minimize the influence of disturbance-induced compliant deformations, i.e., the quasi-static deformations induced by a locally applied force, often modeled using an additional feedthrough term, see, e.g., \cite{fleming2003}. Combined with a moving disturbance source and point-of-interest location, the position dependency of this compliant effect necessitates a position-dependent control approach to obtain the desired performance. This position-dependent controller can be effectively and intuitively realized by combining an observer containing a position-dependent system model with an LTI controller that is robust to the remaining position-dependency, as shown in Figure \ref{fig:fb_obs}.

\section{Conclusions and outlook}
\label{Sect:Conclusions_n_out}

\subsection{Conclusions}
\label{Sect:Conclusions}

This paper provides a general procedure for the identification of position-dependent precision mechatronic systems which consist of a single flexible moving body and with small deformations. This is an essential step for the control of future high-precision motion systems. A key step in the proposed approach is to utilize prior mechanical systems knowledge as embedded in the modal modeling framework to obtain a parsimonious model set. 

In Section \ref{Sect:LTI_ID}, a flexible framework of parametrizations and identification algorithms is proposed that is especially suited for the identification of modal models of mechanical systems. For the considered state-of-the-art industrial wafer stage system, with a total of $7$ considered inputs and $16$ outputs, the proposed identification approach yields a very accurate modal system model with $12$ identified modes. The spline-based interpolation approach proposed in Section \ref{Sect:interp_OAT} provides a robust and effective method to reconstruct the spatial mode shapes, and is successfully applied to reconstruct $9$ of the identified mode shapes. 

Potential applications of the proposed position-dependent modeling approach for control are numerous, including, e.g., the use in global spatio-temporal feedforward control and observer based inferential feedback control. 

\subsection{Outlook}
\label{Sect:Model_interacting_sys}

In this paper, systems are considered that can be written as \eqref{eq:spa_temp_1}--\eqref{eq:spa_temp_2}. Although this description is less general than \eqref{eq:LPV_1}--\eqref{eq:LPV_3}, it is envisaged that the proposed framework of identifying modal models of mechanical systems and subsequently interpolating the spatial system behavior can be extended to be more broadly applicable. In particular, extending the proposed framework to consider the modeling of interacting mechanical subsystems provides the ability to model a variety of relevant mechanical systems. Examples of such interacting mechanical subsystems include the much used H-bridge systems as considered in, e.g., \cite{Steinbuch2003,GrootWassink2005}. By separately considering the spatio-temporal behavior of the subsystems, such as the beam and carriage in an H-bridge system, and modeling the full system behavior as a general interconnection of these component models, a more general class of systems can indeed be described, including systems with position dependent state-matrices. Validating the practical applicability and performance of this approach, as well as the control approaches discussed in Section \ref{Sect:Control_app}, is a subject of ongoing research.

\begin{footnotesize}

\end{footnotesize}

\begin{IEEEbiography}[{\includegraphics[width=1in,height=1.25in,clip,keepaspectratio]{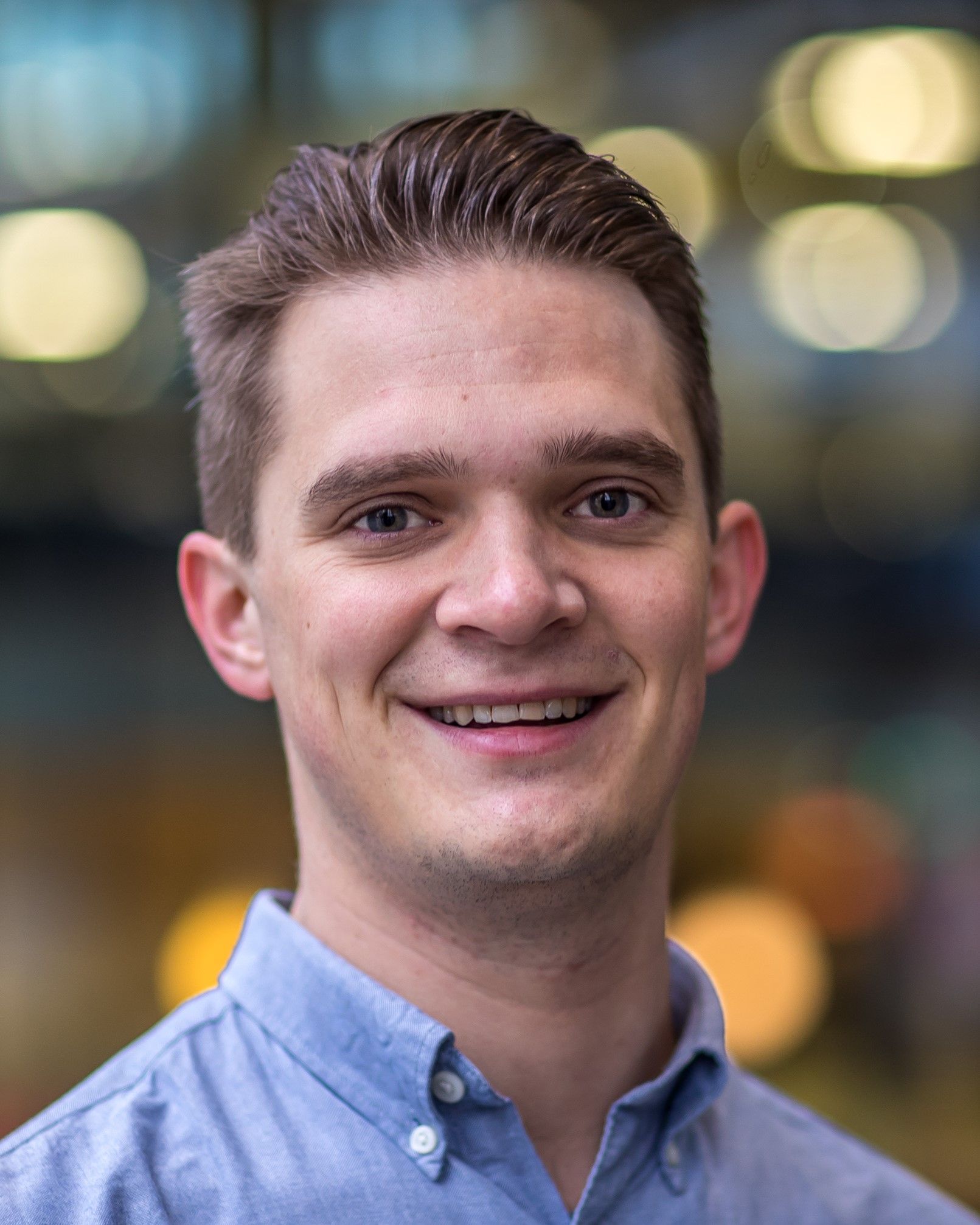}}]%
{Robbert Voorhoeve} received the M.Sc. degree in Mechanical Engineering (cum laude) and Applied Physics in 2013 and his Ph.D. degree in Mechanical Engineering in 2018 from the Eindhoven University of Technology, Eindhoven, The Netherlands. His research interests include system identification, identification for advanced motion control, and control of complex mechatronic systems.
\end{IEEEbiography}

\begin{IEEEbiography}[{\includegraphics[width=1in,height=1.25in,clip,keepaspectratio]{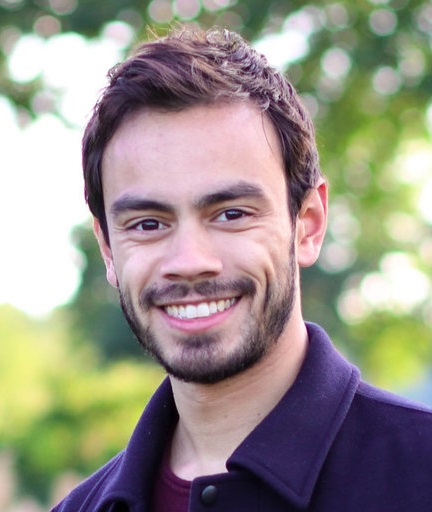}}]%
{Robin de Rozario} received the M.Sc. degree in Mechanical Engineering (cum laude) in 2015 from the Eindhoven University of Technology, Eindhoven, the Netherlands, where he is currently pursuing a Ph.D. degree in the Control Systems Technology group. His research interests include system identification and learning control for high-performance motion systems.
\end{IEEEbiography}

\begin{IEEEbiography}[{\includegraphics[width=1in,height=1.25in,clip,keepaspectratio]{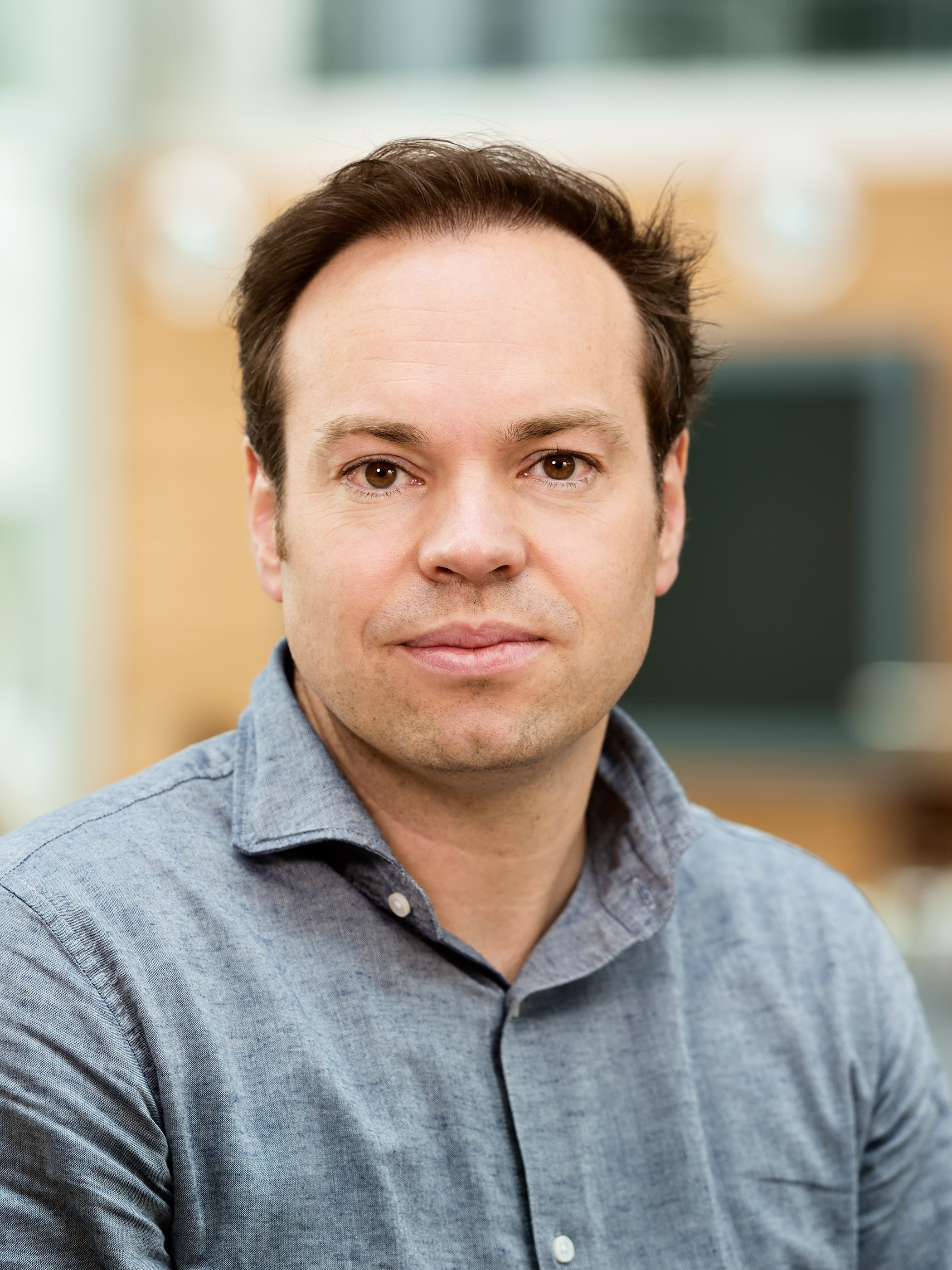}}]%
{Wouter Aangenent} received the M.Sc. degree (cum laude) and the Ph.D. degree in Mechanical Engineering from the Eindhoven University of Technology, Eindhoven, The Netherlands, in 2004 and 2008, respectively. In 2008, he joined the Research Department of ASML, Veldhoven, The Netherlands, where he currently leads the Mechatronics and Control Research Group. 
\end{IEEEbiography}

\begin{IEEEbiography}[{\includegraphics[width=1in,height=1.25in,clip,keepaspectratio]{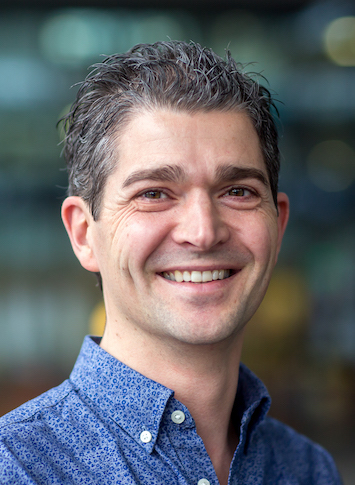}}]%
{Tom Oomen} (SM'06) received the M.Sc. degree (cum laude) and Ph.D. degree from the Eindhoven University of Technology, Eindhoven, The Netherlands. He held visiting positions at KTH, Stockholm, Sweden, and at The University of Newcastle, Australia. Presently, he is associate professor with the Department of Mechanical Engineering at the Eindhoven University of Technology. He is a recipient of the Corus Young Talent Graduation Award, the IFAC 2019 TC 4.2 Mechatronics Young Research Award, the 2015 IEEE Transactions on Control Systems Technology Outstanding Paper Award, the 2017 IFAC Mechatronics Best Paper Award, the 2019 IEEJ Journal of Industry Applications Best Paper Award, and recipient of a Veni and Vidi personal grant. He is Associate Editor of the IEEE Control Systems Letters (L-CSS), IFAC Mechatronics, and IEEE Transactions on Control Systems Technology. He is a member of the Eindhoven Young Academy of Engineering. His research interests are in the field of data-driven modeling, learning, and control, with applications in precision mechatronics.
\end{IEEEbiography}

\end{document}